\documentclass[useAMS,usenatbib]{mnras}%
\usepackage[T1]{fontenc}
\usepackage{ae,aecompl}
\usepackage{times}
\usepackage{amsmath}
\usepackage{amssymb}
\usepackage[english]{babel}
\usepackage[varg]{txfonts}
\usepackage{graphicx}
\usepackage[percent]{overpic}
\usepackage{tikz}
\usetikzlibrary{decorations.pathreplacing,external}
\usepackage{pdfpages}
\usepackage{flafter}
\usepackage{multirow}
\usetikzlibrary{arrows}
\usepackage[]{natbib}
\usepackage{color}

\newcommand{\ud}{d}

\newcommand{\comment}{}

\title[Gravitational Waves]{Gravitational Wave Signals from 3D Neutrino Hydrodynamics
  Simulations of Core-Collapse Supernovae}
\author[H.~Andresen et al.]{
H.~Andresen$^{1,2}$\thanks{E-mail: haakoan@mpa-garching.mpg.de},
B.~M\"uller$^{3,4}$,
E.~M\"uller$^{1}$,
 and H.-Th.~Janka$^{1}$\\
$^{1}$Max-Planck-Institut f\"ur Astrophysik,
   Karl-Schwarzschild-Str.~1, D-85748 Garching, Germany
\\
$^2$Physik Department, Technische Universit\"at M\"unchen, 
James-Franck-Str. 1, 85748 Garching, Germany
\\
$^{3}$Astrophysics Research Centre, School of
   Mathematics and Physics, Queen's University Belfast, Belfast BT7~1NN, United Kingdom\\
$^{4}$Monash Centre for Astrophysics, School of
   Physics and Astronomy, Building 79P, Monash University, Victoria
   3800, Australia
}

\def\LaTeX{L\kern-.36em\raise.3ex\hbox{a}\kern-.15em
    T\kern-.1667em\lower.7ex\hbox{E}\kern-.125emX}
\begin{document}
\maketitle

\begin{abstract}
  We present gravitational wave (GW) signal predictions from
  four 3D multi-group neutrino hydrodynamics simulations of core-collapse
  supernovae of progenitors with $11.2 M_\odot$, $20 M_\odot$, and 
  $27 M_\odot$. GW emission in the pre-explosion phase strongly depends on
  whether the post-shock flow is dominated by the standing accretion
  shock instability (SASI) or convection and differs considerably from
  2D models. SASI activity produces a strong signal component below
  $250 \, \mathrm{Hz}$ through asymmetric mass motions in the gain layer and a non-resonant coupling to the proto-neutron star (PNS). 
  Both convection- and SASI-dominated models show GW emission above $250 \, \mathrm{Hz}$, but with considerably
  lower amplitudes than in 2D. This is due to a different excitation
  mechanism for high-frequency $l=2$ motions in the 
  PNS surface, which are predominantly excited by PNS convection in
  3D. Resonant excitation of high-frequency surface g-modes in 3D by
  mass motions in the gain layer is suppressed compared to 2D because
  of smaller downflow velocities and a lack of high-frequency
  variability in the downflows. In the exploding $20 M_\odot$ model,
  shock revival results in enhanced low-frequency emission
  due to a change of the preferred scale of the convective eddies in
  the PNS convection zone.  Estimates of the expected excess power
  in two frequency bands suggests that second-generation
  detectors will only be able to detect very nearby events, but that
  third-generation detectors could distinguish SASI- and convection-dominated
  models at distances of $\mathord{\sim} 10 \, \mathrm{kpc}$.
\end{abstract}

\section{Introduction}
Despite impressive progress during recent years, the explosion
mechanism powering core-collapse supernovae is still not fully
understood. For ordinary supernovae with explosion energies up to
$\mathord{\sim}10^{51} \,\mathrm{erg}$, the prevailing theory is the
delayed neutrino-driven mechanism (see
\citealp{janka_12,burrows_13} for current reviews). In this scenario,
the shock wave formed during the rebound (bounce) of the inner core
initially stalls and only propagates out to a radius of
$\mathord{\sim}150 \,\mathrm{km}$. The energy needed to revitalise the
shock is provided by the partial re-absorption of neutrinos emitted
from the proto-neutron star (PNS).  

Hydrodynamical instabilities operating behind the stalled shock front
have been found to be crucial for the success of this scenario as they
help to push the shock further out by generating large Reynolds
stresses (or ``turbulent pressure'', see \citealp{burrows_95,murphy_12,couch_15,mueller_15a})
and transporting neutrino-heated material out from the gain radius, which then
allows the material to be exposed to neutrino heating over a longer
``dwell time'' \citep{buras_06b,murphy_08b}. Moreover, if the
instabilities lead to the formation of sufficiently large high-entropy
bubbles, the buoyancy of these bubbles can become high enough to
allow them to rise and expand continuously
\citep{thompson_00,dolence_13,fernandez_15}. 

Two such instabilities have been identified in simulations,
namely the more familiar phenomenon of convection driven by the unstable entropy gradient arising due to neutrino heating
\citep{bethe_90,herant_94,burrows_95,janka_96,mueller_97}, and the
so-called standing accretion shock instability (SASI), which manifests
itself in large-scale sloshing and spiral motions of the shock
\citep{blondin_03,blondin_06,foglizzo_07,ohnishi_06,ohnishi_08,scheck_08,guilet_12,foglizzo_15}.
After initial setbacks in three-dimensional (3D) supernova modelling,
we are now starting to see the emergence of the first generation
of successful 3D simulations of explosions with three-flavour
multi-group neutrino transport, culminating in the recent models
of the Garching and Oak~Ridge groups
\citep{melson_15a,melson_15b,lentz_15} with their rigorous treatment
of the transport and neutrino microphysics in addition to many more
obtained with more approximate transport schemes,
as for example the studies of \citet{takiwaki_12,takiwaki_14}, 
\citet{mueller_15b} and \citet{roberts_16}.\footnote{
\citet{takiwaki_12,takiwaki_14} employ the isotropic diffusion source 
approximation \citep{liebendoerfer_09}
and use further approximations to treat heavy lepton neutrinos.
\cite{takiwaki_14} employ a leakage scheme to account for heavy lepton
neutrinos and \cite{takiwaki_12} neglect the effect of these neutrinos 
altogether. \cite{mueller_15b} utilises the stationary fast multi-group 
transport scheme of \cite{mueller_15a}, which at high optical depths 
solves the Boltzmann equation in a two-stream approximation and
matches the solution to an analytic variable 
Eddington factor closure at low optical depths.
\citet{roberts_16} employ a full 3D two-moment (M1) solver in 
general relativistic simulations, but ignore velocity-dependent
terms.}

Our means to validate these numerical models by observations are
limited. 
Photon-based observations of supernovae and their remnants
(e.g.\ mixing in the envelope, see \citealp{wongwathanarat_15} and references therein; pulsar kicks,
\citealp{scheck_06,wongwathanarat_10b,wongwathanarat_13,nordhaus_12}) provide only
indirect constraints on the workings of the hydrodynamic
instabilities in the inner engine of a core-collapse supernova.
For a nearby, Galactic supernova event,
messengers from the core in the form of neutrinos and gravitational waves (GWs) 
could furnish us with a direct glimpse at the engine. Neutrinos,
for example, could provide a smoking gun for SASI activity through
fast temporal variations \citep{marek_08,lund_10,brandt_11,tamborra_13,tamborra_14b,mueller_14} and
could even allow a time-dependent reconstruction of the shock trajectory
\citep{mueller_14}.

Likewise, a detection of GWs could potentially help to
unveil the multi-dimensional effects operating in the core of a
supernova. The signal from the collapse and bounce of rapidly rotating
iron cores and triaxial instabilities in the early post-bounce phase
has long been studied in 2D (i.e.\ under the assumption of
axisymmetry) and 3D
(e.g.\ \citealp{ott_06_a,dimmelmeier_07_a,dimmelmeier_08,scheidegger_08,abdikamalov_10}). Understanding
the GW signal generated by convection and the SASI in
the more generic case of slow or negligible rotation has proved more
difficult due to a more stochastic nature of the signal. During the
recent years, however, a coherent picture of GW
emission has emerged from parameterised models \citep{murphy_09} and
first-principle simulations of supernova explosions in 2D
\citep{marek_08,mueller_13}: The models typically show an early,
low-frequency signal with typical frequencies of $\mathord{\sim} 100
\,\mathrm{Hz}$ arising from shock oscillations that are seeded by
prompt convection
\citep{marek_08,murphy_09,yakunin_10,mueller_13,yakunin_15}. This
signal component is followed by a high-frequency signal with stochastic
amplitude modulations that is generated by forced oscillatory motions
in the convectively stable neutron star surface layer
\citep{marek_08,murphy_09,mueller_13} with typical frequencies of $300
\ldots 1000 \,\mathrm{Hz}$  that closely trace the
Brunt-V\"ais\"ala frequency in this region \citep{mueller_13}. Prior
to the explosion, these oscillations, tentatively identified as $l=2$
surface g-modes by \citet{mueller_13}, are primarily driven by the
downflows impinging onto the neutron star, whereas PNS
convection takes over as the forcing agent a few hundred
milliseconds after shock revival as accretion dies down.  This
high-frequency contribution dominates the energy spectrum and the
total energy emitted in GWs can reach
$\mathord{\sim} 10^{46} \,\mathrm{erg}$ \citep{mueller_13,yakunin_15}.

Since 3D supernova models have proved fundamentally different to 2D
models in many respects, it stands to reason that much of what we have
learned about GW emission from first-principle 2D
models will need to be revised. In 2D, the inverse turbulent cascade
\citep{kraichnan_76} facilitates the emergence of large-scale flow
structures also in convectively-dominated models and helps to increase
the kinetic energy in turbulent fluid motions in the post-shock region
\citep{hanke_12}. Furthermore, accretion downflows impact the
PNS with much higher velocities in 2D than in 3D \citep{melson_15a} due to the
inverse turbulent cascade and the stronger inhibition of
Kelvin-Helmholtz instabilities at the interface of supersonic
accretion downflows \citep{mueller_15b}. In the SASI-dominated regime,
on the other hand, the additional dimension allows the development of
the spiral mode \citep{blondin_07a,blondin_07b,fernandez_10} in 3D,
which can store more non-radial kinetic energy than pure sloshing
motions in 2D \citep{hanke_13,fernandez_15}, contrary to earlier
findings of \cite{iwakami_08}. Such far-reaching differences
between 2D and 3D cannot fail to have a significant impact on
the GW signal. 

While the impact of 3D effects on the GW signals from
the post-bounce phase has been investigated before, all available
studies have relied on a rather approximate treatment of neutrino
heating and cooling such as simple light-bulb models
\citep{mueller_97,kotake_09,kotake_11}, grey neutrino transport
\citep{fryer_04,mueller_e_12}, or a partial implementation of the
isotropic diffusion source approximation of \citet{liebendoerfer_09}
in the works of \citet{scheidegger_08,scheidegger_10}, which were also
limited to the early post-bounce phase.  Arguably, none of these
previous studies have as yet probed precisely the regimes encountered
by the best current 3D simulations (e.g.\ the emergence
of a strong SASI spiral mode) and therefore cannot be relied upon for quantitative
predictions of GW amplitudes and spectra, which
are extremely sensitive to the nature of hydrodynamic instabilities,
the neutrino heating, and the contraction of the PNS.

In this paper, we present GW waveforms of the
first few hundred milliseconds of the post-bounce phase computed from 3D
models with multi-group neutrino transport. Waveforms have been
analysed for four supernova models of progenitors with zero-age main
sequence (ZAMS) masses of $11.2 M_\odot$, $20 M_\odot$  (for which
we study an exploding and a non-exploding simulation), and $27
M_\odot$. With four simulations based on these three different
progenitors, we cover both the convective
regime ($11.2 M_\odot$) and the SASI-dominated regime ($20 M_\odot$,
$27 M_\odot$). Our aim in studying waveforms from these progenitors is
twofold: On the one hand, we shall attempt to unearth the underlying
hydrodynamical phenomena responsible for the GW emission in different 
regions of the frequency spectrum during different
phases of the evolution.  We shall also compare the GW
emission in 3D and 2D models, which will further illuminate dynamical
differences between 2D and 3D. Furthermore, with 3D models now
at hand, we are in a position to better assess the detectability of
GWs from the post-bounce phase in present and future
instruments than with 2D models affected by the artificial constraint
of axisymmetry.

One of our key findings is that the GW signal 
from SASI-dominated models is clearly differentiated from convection-dominated model 
by strong emission in a low-frequency band around $100 \ldots 200 \, \mathrm{Hz}$. 
Very recently, \cite{kuroda_16} also studied the GW signal 
features (in models using grey neutrino transport) during phases of SASI activity for a $15 M_\odot$ star,
comparing results for three different nuclear equations of state. Going beyond \cite{kuroda_16}, we clarify why this 
signature has not been seen in 2D models and point out that the hydrodynamic processes underlying 
this low-frequency signal are quite complex and seem to require a coupling of SASI motions to deeper 
layers inside the PNS. Moreover, we show that broadband low-frequency GW emission 
can also occur after the onset of the explosion and is therefore not an unambiguous signature of 
the SASI. We also provide a more critical assessment of the detectability of this new signal component, 
suggesting that it may only be detectable with second-generation instruments like Advanced LIGO for a very 
nearby event at a distance of $2 \, \mathrm{kpc}$ or less.

Our paper is structured as follows: We first give a brief
description of the numerical setup and the extraction of
GWs in Section~\ref{sec:numerics}.
In Section~\ref{sec:structure}, we present a short overview
of the GW waveforms and then analyse the hydrodynamical
processes contributing to different parts of the spectrum
in detail. We also compare our results to recent studies
based on 2D first-principle models.
In Section~\ref{sec:obs}, we discuss the detectability
of the predicted GW signal from our
three progenitors by Advanced LIGO \citep{advligo_15},
and by the Einstein Telescope \citep{et_12} as next-generation
instrument. We also comment on possible inferences
from a prospective GW detection. Our
conclusions and a summary of open questions for
future research are presented in Section~\ref{sec:con}.
\section{Simulation setup}
\label{sec:numerics}
\subsection{Numerical Methods}
{\comment The simulations were performed with the \textsc{prometheus-vertex} code \citep{rampp_02,buras_06a}. 
The Newtonian hydrodynamics module
\textsc{prometheus} \citep{mueller_91,fryxell_91} features a dimensionally-split
implementation of the piecewise parabolic method of \cite{colella_84}
in spherical polar coordinates $(r,\theta,\varphi)$. Self-gravity is treated using the monopole
approximation, and the effects of general relativity are accounted for
in an approximate fashion by means of a pseudo-relativistic
effective potential (case~A of \citealt{marek_06}). The neutrino transport
module \textsc{vertex} \citep{rampp_02} solves the energy-dependent two-moment
equations for three neutrino species ($\nu_e$, $\bar{\nu}_e$, and a species $\nu_X$ representing
all heavy flavor neutrinos) using a variable Eddington factor technique.
The ``ray-by-ray-plus'' approximation of \citet{buras_06a} is applied to make the
multi-D transport problem tractable. In the high-density regime, the nuclear equation of  state (EoS) of \citet{lattimer_91} with a bulk incompressibility modulus of nuclear matter of $K=220 \,\mathrm{MeV}$
has been used in all cases.}

\begin{figure*}
\includegraphics[width=0.99\textwidth]{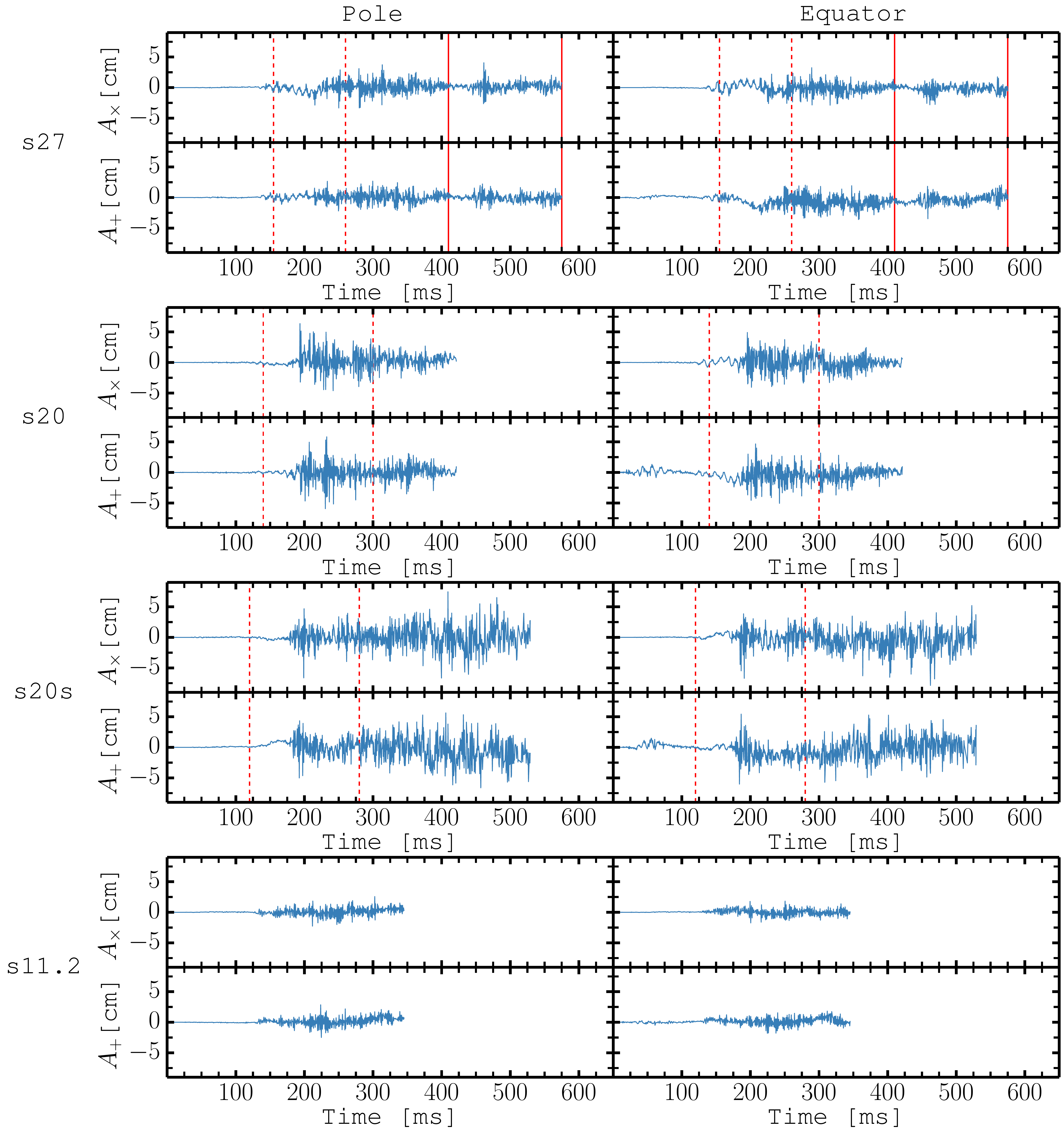}
\caption{GW amplitudes $A_+$ and $A_\times$ as functions of time
  after core bounce.
  From the top: s27, s20, s20s, and s11, respectively. 
  The two columns show the amplitudes for two different viewing angles: an observer
  situated along the $z$-axis (pole; left) and an other observer along the $x$-axis (equator; right) of the computational grid, respectively.
  Episodes of strong SASI activity occur between the vertical red lines; dashed and solid lines are used for
  model s27 to distinguish between two different SASI episodes.
\label{fig:amps}}
\end{figure*}

\subsection{Supernova Models}
\subsubsection{\comment 3D  Models}
We study four 3D models based on three solar-metallicity progenitor stars with ZAMS masses of $11.2 M_\odot$ \citep{woosley_02},$20 M_\odot$ \citep{woosley_07} and $27 M_\odot$ \citep{woosley_02}. 
{\comment An initial grid resolution of $400\times 88 \times 176$ zones in $r$, $\theta$, and $\varphi$ was used for the 3D models, and more radial grid zones were added
during the simulations to maintain sufficient resolution around the PNS surface.}
{\comment The innermost $10 \, \mathrm{km}$ were simulated in spherical symmetry to avoid
excessive limitations on the time step when applying a spherical polar grid.}

\begin{itemize}
\item \textbf{s11.2:} 
Model s11.2 \citep{tamborra_14a} is based on the solar-metallicity $11.2 M_\odot$ progenitor of \citet{woosley_02}. 
This model exhibits transient shock expansion after
the infall of the Si/O shell interface, but falls slightly short of
an explosive runaway.  After the average shock radius reaches
a maximum of $\mathord{\approx} 250 \, \mathrm{km}$ at a time
of $\mathord{\approx} 200 \, \mathrm{ms}$ after bounce, the shock recedes 
and shock revival is not achieved by the end of the simulation $352 \, \mathrm{ms}$ after core bounce.
The post-shock region is dominated by buoyancy-driven convection; because of the large shock radius
no growth of the SASI is observed. The convective bubbles remain of moderate
scale: Even during the phase of strongest shock expansion around
$\mathord{\sim} 200 \, \mathrm{ms}$ after bounce when the shock deformation
is most pronounced and the kinetic energy in convection motions reaches
its peak value, the bubbles subtend
angles of no more than $\lesssim 60^\circ$.
\item \textbf{s20:}
Model s20 is based on the $20 M_{\odot}$  solar-metallicity progenitor
of \citet{woosley_07} and has been discussed in greater
detail in \citet{tamborra_13,tamborra_14b}, where quasi-periodic modulations of the neutrino
emission were analysed and traced back to SASI-induced variations of the mass-accretion flow to the PNS. 
No explosion is observed by the end of the simulation 421 ms post bounce.
There is an extended phase of strong SASI activity (dominated by the spiral mode) 
between 120 and $280 \, \mathrm{ms}$ after core bounce. After a period
of transient shock expansion following the infall of the Si/O shell
interface, SASI activity continues,
but the kinetic energy in the SASI remains considerably
smaller than during its peak between $200$ and $250 \, \mathrm{ms}$.
\item \textbf{s20s:} This model is based on the same $20 M_\odot$
progenitor as s20, but a non-zero contribution from
strange quarks to the axial-vector coupling constant,
$g_\mathrm{a}^\mathrm{s}=-0.2$, from neutral-current neutrino-nucleon scattering was assumed \citep{melson_15b}.
This modification of the neutrino interaction rates results in a successful
explosion \citep{melson_15b}. Shock revival sets in around $300 \, \mathrm{ms}$ after bounce.
Prior to shock revival, the post-shock
flow is dominated by large-scale SASI sloshing motions between
$120$ and $280 \, \mathrm{ms}$ post-bounce. By the end of the
simulation $528 \, \mathrm{ms}$ post-bounce, the average shock
radius is $\mathord{\approx} 1000 \, \mathrm{km}$, and a strong global
asymmetry stemming from earlier SASI activity remains imprinted onto the post-shock flow. Asymmetric accretion onto the PNS still continues, but the accretion rate is reduced by a factor of $\mathord{\sim} 2$
compared to model s20.
\item \textbf{s27:} Our most massive model is based on the 
$27 M_\odot$ solar-metallicity progenitor of \citet{woosley_02} and has been
discussed in greater detail in \citet{hanke_13} and, for SASI-induced neutrino emission variations, by \citet{tamborra_13,tamborra_14b}.
Shock revival did not occur by the end of the simulation $575 \, \mathrm{ms}$ after
bounce. There are two episodes of pronounced SASI activity that are
interrupted by a phase of transient shock expansion following the
infall of the Si/O interface. The first SASI phase takes place
between $120$ and $260 \, \mathrm{ms}$ post-bounce and the second
period sets in around $410 \, \mathrm{ms}$ post-bounce and lasts until
the end of the simulation. 
\end{itemize}

\subsubsection{\comment 2D  Models}

{\comment
In addition to the 3D models, we also analyse two 2D models based on the same progenitor
as s27. 
\begin{itemize}
\item \textbf{s27-2D:} Model s27-2D was simulated with the same numerical setup as s27 (see \citealp{hanke_13}),
with an initial grid resolution of $400 \times 88$ zones in $r$ and $\theta$ and the 
innermost $10 \, \mathrm{km}$ being simulated in spherical symmetry to allow for
optimal comparison with the 3D model.
SASI activity sets in about $150 \, \mathrm{ms}$ after
core bounce. Between $220 \, \mathrm{ms}$ and $240 \, \mathrm{ms}$ after bounce the accretion rate drops significantly
after the Si/O shell interface has crossed the shock.
The decreasing accretion rate leads to shock expansion, and
shock revival occurs around $300 \, \mathrm{ms}$ post bounce.
\end{itemize}
\begin{itemize}
\item \textbf{G27-2D:} In order to compare our results to those of a 
relativistic 2D simulation of the SASI-dominated s27 model, 
we also reanalyse the 2D model G27-2D presented
by \cite{mueller_13}, which was simulated with \textsc{coconut-vertex} \citep{mueller_10}. 
\textsc{coconut} \citep{dimmelmeier_02_a,dimmelmeier_04} uses a
directionally-unsplit implementation of the piecewise parabolic
method (with an approximate Riemann solver) for general relativistic hydrodynamics in spherical polar coordinates. 
The metric equations are solved in the extended conformal flatness approximation \citep{cordero_09}. 
The model was simulated with an initial grid resolution of $400 \times 128$ zones in $r$ and $\theta$, 
with the innermost $1.6 \, \mathrm{km}$ being simulated in spherical symmetry to reduce time step limitations.
For consistency, we recompute the gravitational wave amplitudes for this model based on the relativistic stress formula 
(Appendix~A of \citealp{mueller_13}) instead of the time-integrated quadrupole formula with centred differences as originally used by \cite{mueller_13}. 
The stress formula leads to somewhat larger amplitudes particularly at late times when central differencing
is no longer fully adequate due to the increasing signal frequency.

The model is characterised by strong post-shock convection for the first $50 \, \mathrm{ms}$
after core bounce, which is followed by a phase of
strong SASI activity. Around $120 \, \mathrm{ms}$ after core bounce the average
shock radius starts to increase steadily. The criterion for runaway shock expansion is met
approximately $180 \, \mathrm{ms}$ after bounce and 
the shock is successfully revived at $\mathord{\sim 210} \, \mathrm{ms}$ post bounce.

{\comment The evolution of models G27-2D and s27-2D differs
  significantly during the pre-explosion phase: In G27-2D large-scale
  deformation of the shock already occurs $\mathord{\sim} 50$ ms after
  bounce, without a preceding phase of hot-bubble convection
  \citep{mueller_13}.  s27-2D develops SASI activity later: Since the
  average shock radius is $\sim$20-30 km larger in model s27-2D than
  in model G27-2D, conditions favour the development of
  neutrino-driven convection. Consequently, s27-2D shows an initial
  phase of convection before SASI activity sets in when the shock
  starts to retract $\sim$100-150 ms after bounce.  Due to the early
  development of SASI activity in model G27-2D at a time when the
  accretion rate is high, particularly strong supersonic downflows
  onto the PNS develop.

A possible reason for stronger and earlier SASI activity in model
G27-2D is that, in contrast to model s27-2D, model G27-2D exhibits a
phase of strong \emph{prompt} post-shock convection (between a few ms
after bounce and about $50 \, \mathrm{ms}$ after bounce), which leaves
the shock appreciably deformed with $|a_1|/a_0 \sim 0.01-0.02$ as
shown in Fig.~7 of \citet{mueller_12b}. Therefore the SASI amplitude
only needs to grow by a factor of $\mathord{\sim} 30$ to reach the
non-linear regime. In \citet[][Fig.~2]{hanke_13} the $l = 1$ amplitude
is much smaller at early times.  Such differences in the post-bounce
evolution can have a variety of reasons. Besides the pure
stochasticity of simulations, the initial perturbations may also play
a role: Model G27-2D was simulated in 2D from the onset of core
collapse, while model s27-2D was started from a spherical model with
seed perturbations imposed $15 \, \mathrm{ms}$ after core bounce. The
presence or absence of strong prompt post-shock convection also
depends on the details of the entropy and electron fraction profiles,
which are determined by the exact shock dynamics during the first
milliseconds after core bounce. Without a very careful analysis of all
the differences between the two simulations, we are not able to
localise the origin of the differences between model G27-2D and model
s27-2D in the different gravity treatment or any of the other
aforementioned aspects.

Despite (or because of) the differences in the dynamics of the post-shock flow to
  s27-2D, model G27-2D is useful for illustrating the differences of
  the 3D model to the extreme end of the spectrum of recent 2D models
  in terms of peak GW amplitude and illustrates the mechanism of GW
  emission by stochastic surface g-mode excitation due to overshooting
  plumes from the gain region \citep{marek_08,murphy_09,mueller_13} in
  the clearest form. 
 }

\end{itemize}
}
\section{Structure and Origin of the Gravitational Wave Signal}
\label{sec:structure}
{\comment
The different 3D models used in our analysis probe distinctly
different regimes that can be encountered in supernova cores.
In this section, we will investigate how these dynamical differences are reflected
in the GW signals. We also compare our 3D models to the two 2D models and investigate
how and why the GW signal changes when going from 2D to 3D.}
\subsection{Gravitational Wave Extraction}
In order to extract the GW signal from
the hydrodynamical simulations, we post-process our simulations using
the quadrupole stress formula \citep{finn_89,nakamura_89,blanchet_90}.  Here,
we only give a concise description of the formalism and refer the reader to 
\cite{mueller_e_12}\footnote{Note, however, that the description of the formalism in
\cite{mueller_e_12} contains some typos: Their Eq.~(24) is incomplete. The superscript TT is missing from $\ddot{Q}_{ij}$, as is also the case in Eq.~(22) and (23),
and, more importantly, the trace term is missing.} for a full explanation.

In the transverse traceless (TT) gauge and the far-field limit the metric perturbation, $\bmath{h^\mathrm{TT}}$, 
can be expressed in terms of the amplitudes of the two independent polarisation modes in the following way,
\begin{equation}
\bmath{h}^\mathrm{TT}(\bmath{X},t) = \frac{1}{D}   \left [ A_{+} \bmath{e}_{+} + A_{\times} \bmath{e}_{\times} \right ].
\end{equation}
Here, $D$ denotes the distance between the source and the observer, $A_+$ denotes the wave amplitude of the plus-polarised mode, $A_\times$ is the wave amplitude of the cross-polarised mode and
$\mathbf{e}_{\times}$ and $\mathbf{e}_{+}$ denote the unit polarisation tensors. The unit polarisation tensors are given by
\begin{equation}
\bmath{e}_{+}  = \bmath{e}_{\theta} \otimes \bmath{e}_{\theta} - \bmath{e}_{\phi} \otimes \bmath{e}_{\phi},
\end{equation}
\begin{equation}
\bmath{e_{\times}} = \bmath{e}_{\theta} \otimes \bmath{e}_{\phi} + \bmath{e}_{\phi} \otimes \bmath{e}_{\theta},
\end{equation}
where $\bmath{e}_{\theta}$ and $\bmath{e}_{\phi}$ are the unit vectors in the $\theta$ and $\phi$
direction of a spherical coordinate system and $\otimes$ denotes the tensor product.
Using the quadrupole approximation in the slow-motion limit, the amplitudes $A_{\times}$ and $A_{+}$ can be computed from the second time derivative 
of the symmetric trace-free (STF) part of the mass quadrupole tensor $Q$~\citep{oohara_97},
\begin{equation}
\label{eq:aplus}
A_{+} = \ddot{Q}_{\theta \theta} - \ddot{Q}_{\phi \phi},
\end{equation}
\begin{equation}
\label{eq:ax}
A_{\times} =2 \ddot{Q}_{\theta \phi}.
\end{equation}
The components of $Q$ in the orthonormal basis associated with
spherical polar coordinates used in this formula can be obtained from
the Cartesian components $\ddot{Q}_{ij}$ of $\ddot{Q}$ \citep{oohara_97,nakamura_87}.  Using the
continuity and momentum equations to eliminate time derivatives
\citep{oohara_97,finn_89,blanchet_90}, the Cartesian components can be obtained as:
\begin{equation} \label{eq:STFQ}
\ddot{Q}_{ij} =\mathrm{STF} \left [2 \frac{G}{c^4} \int \ud^3 x \, \rho \left ( v_i v_j - x_i \partial_j \Phi \right ) \right].
\end{equation}
Here, $G$ is Newton's gravitational constant, $c$ is the speed of
light, and $v_i$ and $x_i$ are the Cartesian velocity components and
coordinates ($i = 1,2,3$), respectively. The gravitational potential
$\Phi$ is the gravitational potential used in the simulations (with post-Newtonian
corrections taken into account). $\mathrm{STF}$ denotes the projection operator
onto the symmetric trace-free part.
The advantage of this form is that
the second-order time derivatives are transformed into first-order
spatial derivatives, thus circumventing problems associated with the
numerical evaluation of second-order time derivatives. 
Using standard
coordinate transformations between Cartesian and spherical
coordinates, we obtain \citep{oohara_97,nakamura_87} the components 
$\ddot{Q}_{\theta \theta}$, $\ddot{Q}_{\phi \phi}$, and
$\ddot{Q}_{\theta \phi}$ needed in Eq.~(\ref{eq:aplus}) and (\ref{eq:ax}),
\begin{eqnarray}
\label{eq:qtp}
\ddot{Q}_{\theta \phi} =&  \left (\ddot{Q}_{22} - \ddot{Q}_{11} \right ) \cos{\theta}\sin{\phi}\cos{\phi} \nonumber \\
&+ \ddot{Q}_{12} \cos{\theta} \left (\cos^2 \phi - \sin^2 \phi \right ) \nonumber \\ 
&+ \ddot{Q}_{13} \sin \theta \sin \phi - \ddot{Q}_{23} \sin \theta \cos\phi,
\end{eqnarray}
\begin{eqnarray}
\ddot{Q}_{\phi \phi} &= \ddot{Q}_{11} \sin^2 \phi + \ddot{Q}_{22} \cos^2 \phi - 2 \ddot{Q}_{12} \sin{\phi}\cos{\phi}
\end{eqnarray}
and
\begin{eqnarray}
\ddot{Q}_{\theta \theta} &= \left ( \ddot{Q}_{11} \cos^2 \phi + \ddot{Q}_{22} \sin^2 \phi +  2 \ddot{Q}_{12} \sin{\phi} \cos{\phi} \right) \cos^2 \theta \nonumber \\
&+ \ddot{Q}_{33} \sin^2 \theta - 2 \left (\ddot{Q}_{13} \cos{\phi} + \ddot{Q}_{23} \sin{\phi} \right ) \sin{\theta} \cos{\theta}. 
\end{eqnarray}

In axisymmetry the only independent component of $\bmath{h}^\mathrm{TT}$ is 
\begin{equation}
h^\mathrm{TT}_{\theta \theta} = \frac{1}{8}\sqrt{\frac{15}{\pi}} \sin^2{\theta} \frac{A_{20}^\mathrm{E2}}{D},
\end{equation}
where $D$ is the distance to the source, $\theta$ is the inclination angle of the observer with respect to the
axis of symmetry, and $A_{20}^\mathrm{E2}$ represents the only non-zero quadrupole amplitude.
In spherical coordinates $A_{20}^\mathrm{E2}$ can be expressed as follows
\begin{eqnarray} \label{eq:2dquad}
A_{20}^\mathrm{E2} (t) =  \frac{G}{c^4} \frac{16 \pi^{3/2}}{\sqrt{15}} \int_{-1}^{1}\int^{\infty}_0 \rho \left [ v_r^2(3 z^2 - 1)+ \right. \nonumber \\
v_{\theta}^2(2-3 z^2) - v_{\phi}^2 - 6 v_r v_{\theta} z\sqrt{1-z^2} + \nonumber \\
r \partial_r \Phi (3 z^2 - 1) +\left. 3 \partial_{\theta}z\sqrt{1-z^2} \Phi \right ]r^2 dr \, dz.
\end{eqnarray}
Here, $v_i$ and $\partial_i$ ($i = r, \theta, \phi$) represent the velocity components and derivatives, respectively, along
the basis vectors of the spherical coordinate system, and $z \equiv \cos \theta$.
For details we refer the reader to \cite{mueller_97}.

In this work, we disregard the contribution of anisotropic neutrino
emission \citep{epstein_78} to the GW signal. Due
to its low-frequency nature, it is of minor relevance for the
detectability and does not affect the waveforms appreciably
in the frequency range $\gtrsim 50 \, \mathrm{Hz}$ that is of
primary interest to us in this work.

\subsection{Overview of Waveforms}
\subsubsection{Waveforms}
\label{sec:waveforms}
Amplitudes for GWs generated by asymmetric mass motions are shown in
Fig.~\ref{fig:amps}. For each progenitor, we show two panels
representing the cross and plus polarisation for two different
observer positions. The two columns show the amplitudes for two 
different viewing angles, the right and left column representing
observers situated along the z-axis (pole) and x-axis (equator) of the computational grid, respectively.
\footnote{In our post-processing we chose to sample the GW signal at observer directions corresponding to 
cell centres of the simulation grid and as a consequence the two directions do not exactly correspond to the
north pole ($\theta = 0, \phi = 0$) and the equator ($\theta = \pi, \phi = 0$), but are offset by half of the angular resolution.
Hench, the coordinates of the polar and equatorial observer become ($\pi/176,\pi/176$) 
and ($\pi - \pi/176,\pi/176$), respectively.}
Since our (nonrotating) models do not exhibit a signal from a rotational bounce,
and since (EoS-dependent) prompt post-shock convection is weak, 
the waveforms exhibit an initial quiescent phase. This is followed by a rather 
stochastic phase with amplitudes of several centimetres once convection or the SASI have fully developed. 
The correlation of stronger GW emission
with the onset of strong, non-linear SASI activity
in model s20, s20s, and s27 is illustrated
by dashed and solid lines bracketing phases of particularly
violent SASI oscillations.

The signal from early SASI activity triggered by prompt convection a
few tens of milliseconds after bounce, which is typically rather
prominent in 2D
\citep{marek_08,murphy_08,yakunin_10,mueller_13,yakunin_15},
is thus strongly reduced in 3D. It is only clearly visible in
the waveforms of s20 and s20s, while s11.2 and s27 only show traces
of this component in some directions. 
The stochastic modulation of the later signal is reminiscent
of 2D models, but the amplitudes are significantly lower ($\mathord{\lesssim} 4
\, \mathrm{cm}$) compared to several tens of $\mathrm{cm}$ in first-principle 2D
models \citep{marek_08,yakunin_10,mueller_13,yakunin_15}. The
reduction in 3D is far stronger than could be expected from a mere
projection effect (in agreement with parameterised models of
\citealt{mueller_e_12}).

Prior to a post-bounce time of $\mathord{\sim} 200 \, \mathrm{ms}$,
the waveforms for the three SASI-dominated models
are clearly dominated by a low-frequency signal
(in very much the same fashion as for early SASI activity
a few tens of milliseconds after bounce in 2D). This already indicates
that the relative importance of the low- and high-frequency
components of the signal during the accretion phase is different
in 3D compared to 2D,  where low-frequency
emission (triggered by prompt convection) only dominates for the first tens
of milliseconds. 

In the exploding model s20s with $g_\mathrm{a}^\mathrm{s}=-0.2$, we
observe a tendency towards somewhat higher peak amplitudes than during
the accretion phase well after the onset of the explosion
($\mathord{\sim} 300 \, \mathrm{ms}$). This tendency is, however, much
less pronounced than in 2D models. The monotonous ``tail'' in the
matter signal from anisotropic shock expansion
\citep{murphy_09,yakunin_10} is noticeably absent, although no undue
importance should be attached to this because it may take several
hundreds of milliseconds for the tail to develop \citep{mueller_13}.

\subsubsection{Energy Spectra}
\begin{figure}
\includegraphics[width=0.99\linewidth]{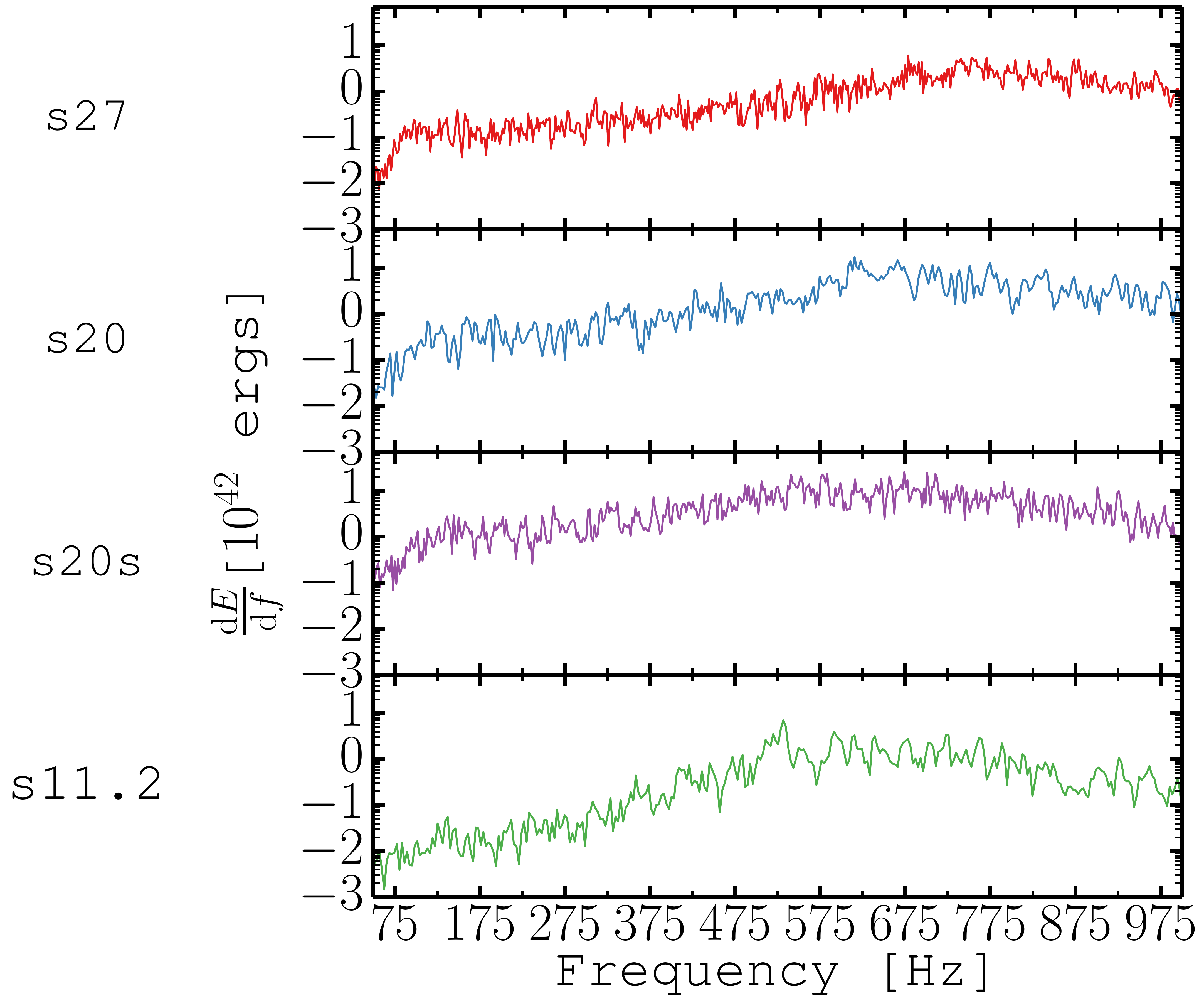}
\caption{Time-integrated GW energy spectra $\ud E/ \ud f$ for models s27,
s20, s11.2, and s20s (top to bottom). The spectra
are computed from the Fourier transform of the entire waveform
without applying a window function. The y-axis is given in a logarithmic scale.
\label{fig:energy_spectra}}
\end{figure}

Time-integrated energy spectra for each of the models are shown
in Fig.~\ref{fig:energy_spectra}. These are computed from
the Cartesian components of the mass quadrupole tensor as
\begin{eqnarray}
\frac{\ud E}{\ud f}=
\frac{2 c^3}{5 G} (2\pi f)^2\left[|\widetilde{\ddot{Q}}_{xx}|^2+\widetilde{|\ddot{Q}}_{yy}|^2+|\widetilde{\ddot{Q}}_{zz}|^2 \right. \nonumber \\ 
\left.+2\left ( |\widetilde{\ddot{Q}}_{xy}|^2 + |\widetilde{\ddot{Q}}_{xz}|^2+|\widetilde{\ddot{Q}}_{yz}|^2 \right )\right],
\end{eqnarray}
where tildes denote Fourier transforms, and $f$ is the
frequency. We define the Fourier transform as follows:
\begin{equation}
\widetilde{\ddot{Q}}_{ij}(f) = \int_{-\infty}^{\infty} \ddot{Q}_{ij}(t) e^{-2 \pi i f t} \ud t. 
\end{equation}
The energy spectra of the three SASI-dominated models are relatively flat. This is significantly
different from 2D models, where the energy spectra
are dominated by a peak at several hundreds of $\mathrm{Hz}$
\citep{marek_08,mueller_13,yakunin_15}. Model s11.2, on the other hand, 
more closely resembles the 2D energy spectra, although
the total energy emitted in GWs is considerably lower than in typical 2D models. 
In addition the peak values of $\ud E/\ud f$
are considerably higher in the SASI models than for s11.2.

\subsubsection{The Signal in the Time-Frequency Domain}
In order to dissect the signal further, we apply a short-time
Fourier transform (STFT) to our waveforms. For a discrete time series
the STFT is obtained by applying the discrete Fourier transform (DFT)
to the signal with a sliding window.
In this work, we define the DFT, $\widetilde{X}_k$, as follows: 
\begin{equation} \label{eq:DFT}
\widetilde{X}_k (f_k) = \frac{1}{M}  \sum^M_{m=1} x_m e^{-2\pi i k m/N},
\end{equation}
Here, $x_m$ is the time series obtained by sampling the underlying continuous signal at $M$ discrete times. 
$f_k = k/T$ is the frequency of bin $k$, where $T$ is the duration of the signal.

\begin{figure*}
\includegraphics[width=0.99\textwidth]{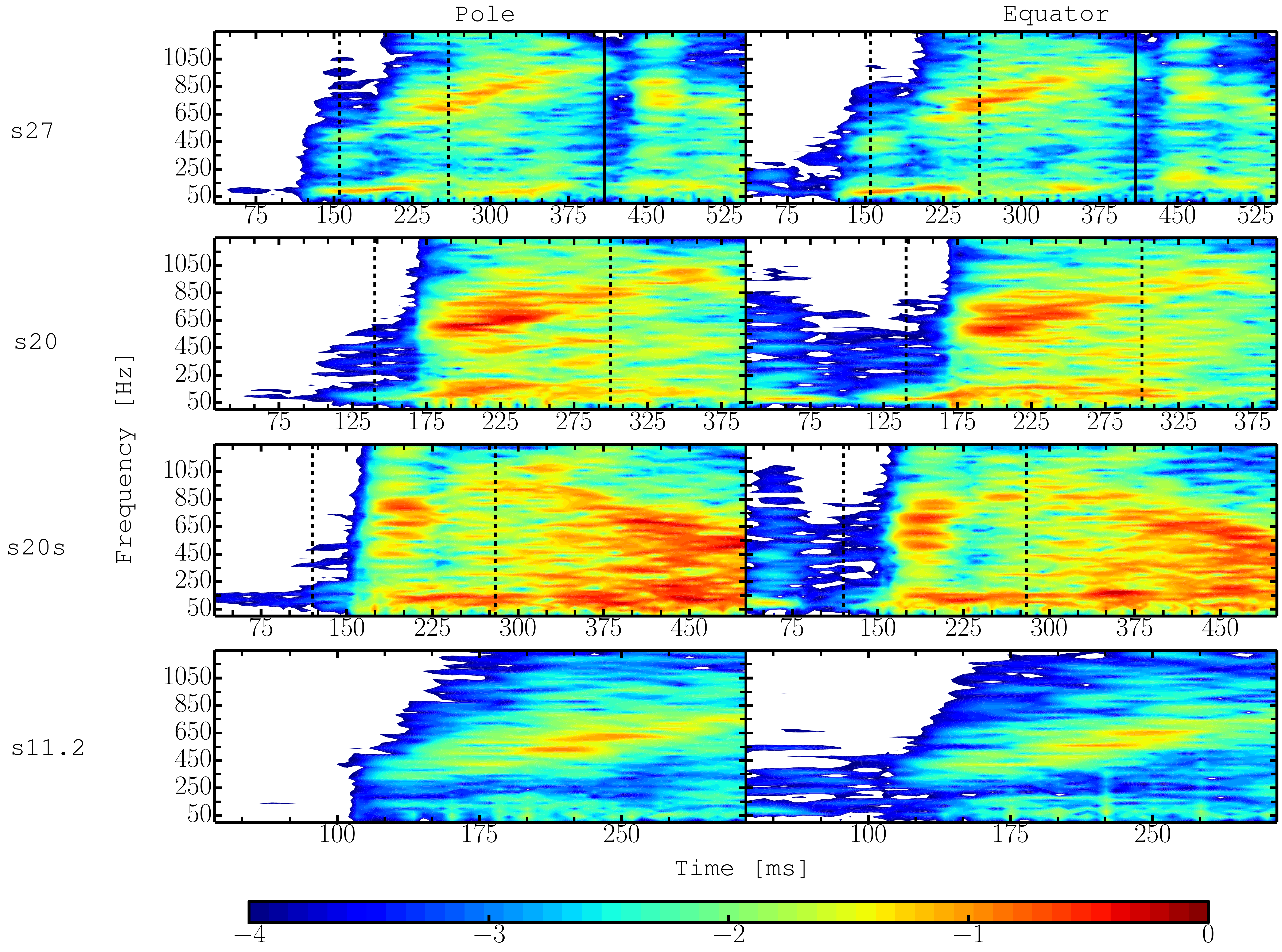} \\
\caption{Amplitude spectrograms for a sliding window of 50 ms and two different observer
  directions, summed over the two polarisation modes 
  ($|\widetilde{A}_+|^2 + |\widetilde{A}_\times|^2$). The
  different rows show the results for models s27, s20, s20s, and s11.2. (top to bottom).
  The two columns shows the spectrograms for two different viewing angles, the right and left column represent
  observers situated along the z-axis (pole) and x-axis (equator) of the computational grid, respectively.
  The time is given in ms after core bounce. Vertical lines bracket SASI episodes. All panels have been normalised by the same global factor.
  The colour bar is given in a logarithmic scale.
\label{fig:spectrograms}}
\end{figure*}

The resulting amplitude spectrograms for a sliding window of $50 \, \mathrm{ms}$ are shown in
Fig.~\ref{fig:spectrograms}. The spectrograms show the sum of the squared Fourier
components of the cross and plus polarisation modes,
$|\widetilde{A_+}|^2 + |\widetilde{A_{\times}}|^2$. Before applying the
DFT we convolve the signal with a Kaiser window with shape parameter $\beta = 2.5$. Frequencies
below $50 \, \mathrm{Hz}$ and above $1100  \, \mathrm{Hz}$ are filtered out of the resulting DFT. The amplitude spectrograms are computed
for the same two observer directions as before.   

All of the models exhibit the distinct high-frequency (here defined to be emission at
frequencies greater than 250 Hz) component familiar from 2D models with a slow, secular increase in the peak frequency.

The SASI-dominated models stand apart from model s11.2 in that they show an additional low-frequency component
(below $250 \, \mathrm{Hz}$) at late times (i.e.\ \emph{not} associated with prompt
convection). No such distinct low-frequency emission has been observed in
spectrograms from 2D models \citep{murphy_09,mueller_13}. The
low-frequency component is clearly separated from the high-frequency
emission by a ``quiet zone'' in the spectrograms. The frequency
structure of the low-frequency component is rather complicated, and
especially for models s20 and s20s it is rather
broad-banded. There is also a directional dependence as can be seen, for example,
from the later onset of low-frequency emission in the polar direction compared to the equatorial direction, 
in model s20 (second row in Fig.~\ref{fig:spectrograms}).

During the explosion phase, we find increased power in the high-frequency band corresponding to the increased peak amplitudes discussed
in Section~\ref{sec:waveforms}. However, the most conspicuous change after the onset
of the explosion consists in a considerable increase
of broadband power at low frequencies.
Close inspection of Fig.~\ref{fig:amps} shows that
the enhanced low-frequency emission can also be seen
directly in the amplitudes: The amplitude
``band'' defined by stochastic high-frequency oscillations
is clearly not centred at zero amplitude, but
exhibits a significant low-frequency modulation.

Typical frequencies of the order of $100 \, \mathrm{Hz}$ as well as a
vague temporal correlation of the low-frequency emission with periods
of strong sloshing/spiral motions suggest a connection with
SASI activity. However, model s27 (top row in
Fig.~\ref{fig:spectrograms}) also shows low-frequency emission during
the phase between $280 \, \mathrm{ms}$ and $350 \, \mathrm{ms}$ after
bounce when the SASI is relatively quiet. If the signal were
directly due to the SASI, one would expect the phases of strong SASI
and strong low-frequency emission to coincide. There may also be
correlations between the low- and high-frequency emission as suggested
by the fact that model s20 with the strongest
low-frequency emission also exhibits the strongest high-frequency
signal. Moreover, the source of enhanced low-frequency emission after
shock revival is not immediately intuitive since the SASI no longer
operates during this phase.  This calls for a closer investigation of
the hydrodynamic processes responsible for the emission of the two
signal components.

\subsection{Spatial Location of Underlying Hydrodynamical Instabilities} \label{sec:spaceloc}
Which regions of the simulation volume contribute to the different GW
components? The emission of GWs cannot be strictly
localised, but one can nonetheless still partition the computational volume 
in the quadrupole formula (\ref{eq:STFQ}) into different regions,
and consider the formal contributions of each of these to the total
signal. While this may not amount to a strict localisation
of GW emission as coming from  a specific region, such a partitioning
nevertheless helps to detect fluid motions with the required
temporal and frequency structure to account for different components
of the signal {\comment in conjunction with the temporal
evolution of the amplitudes and the spectral power. This procedure
cannot replace a more rigorous identification of GW-emitting modes,
which must, however, be left to the future.}

\begin{figure}
\includegraphics[width=0.99\linewidth]{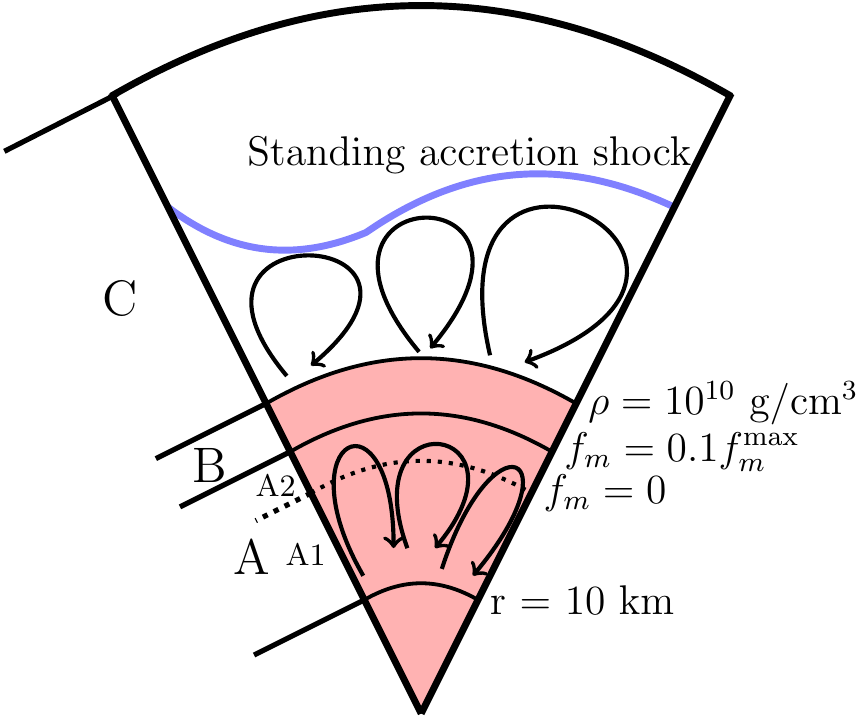}
\caption{Schematic overview of the regions of hydrodynamical activity. In
  Section~\ref{sec:spaceloc} we investigate the contribution to
  the total GW signal from three different layers. The PNS, indicated by
  the shaded red area, is divided into two layers: Layer~A includes
  the convectively unstable region in the PNS (layer~A1) and the overshooting
  layer~A2 directly above it. The boundary between the convective
  layer and the overshooting layer is indicated by a dashed curve
  within layer~A.
  The second layer, layer~B, extends from the
  top of the overshooting region and out to the PNS surface, defined
  by a fiducial density of $10^{10} \, \mathrm{g} \, \mathrm{cm}^{-3}$. Layer~C
  extends from the PNS surface to the outer boundary of our simulation
  volume. Layer~C therefore includes the post-shock region, the
  standing accretion shock (indicated by the blue line), and the
  pre-shock region. Formal definitions of the boundaries between
  layers are given on the right hand side, see
  Section~\ref{sec:spaceloc} for details.
\label{fig:PNSski}}
\end{figure}

In this work, we divide the integration volume into three layers~A, B, and C (see
Fig.~\ref{fig:PNSski}). The PNS is split into two layers, the
``convective layer'' (layer~A) and the convectively stable ``surface layer''(layer~B).  The convectively stable inner core ($r<10 \, \mathrm{km}$) is not considered in our analysis because it is simulated in spherical symmetry and consequently does not contribute to the GW emission.
A third layer (layer~C) comprises the region between the outer boundary
of the PNS (defined by a density of $10^{10} \, \mathrm{g}
\, \mathrm{cm}^{-3}$) and the outer boundary of the grid. We refer
to this region as ``post-shock'' region because
only  motions in the post-shock region and the deceleration of matter
at the shock effectively contribute to the signal from this layer.

\begin{figure}                                       
\includegraphics[width=0.49\textwidth]{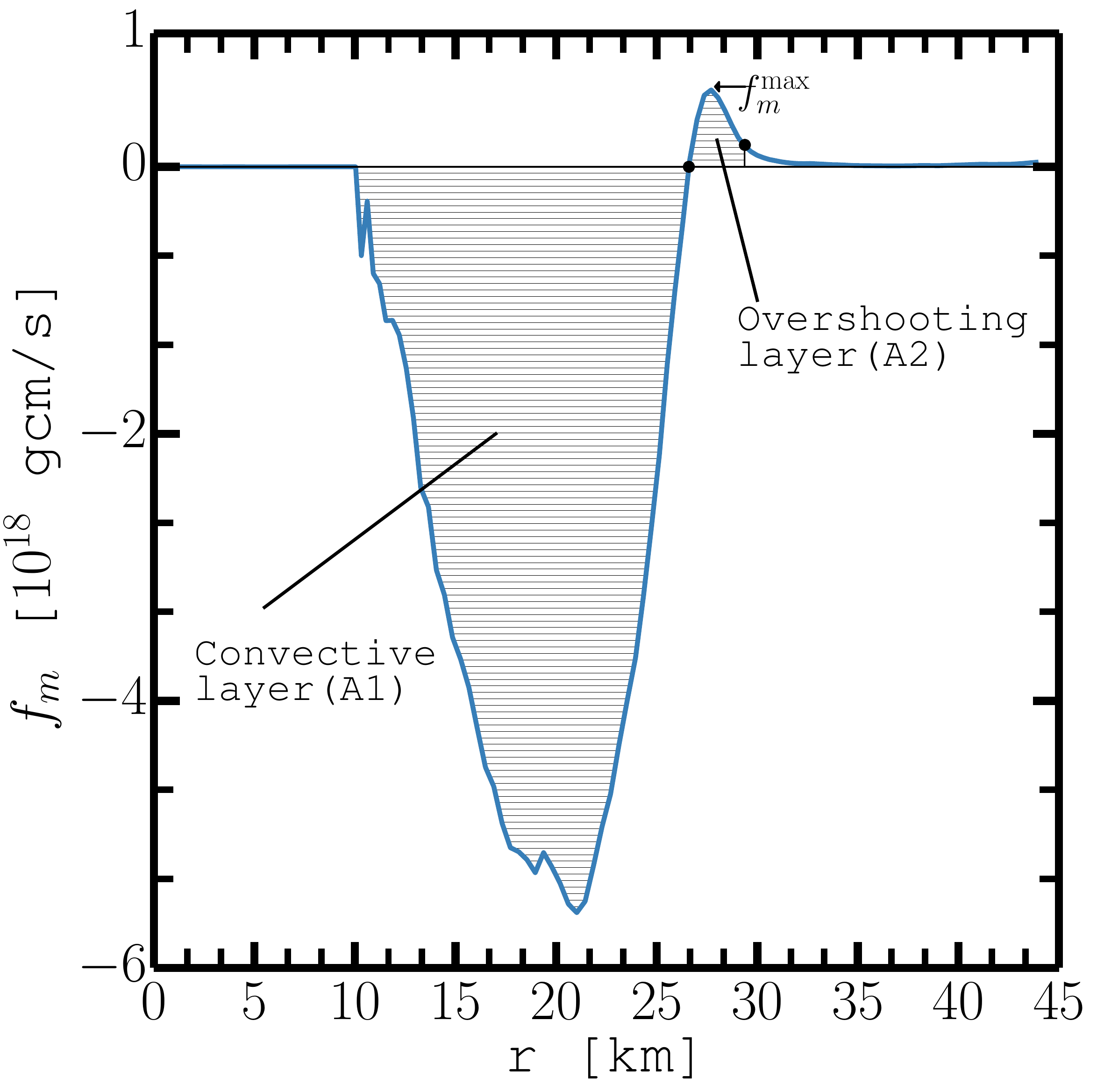}
\caption{Turbulent mass flux $f_m$ (blue curve) for model s27, calculated $192 \, \mathrm{ms}$ after core bounce.
The shaded region indicates the convectively unstable region and the overshooting layer, which
are lumped together as layer~A  (see Fig.~\ref{fig:PNSski}).
\label{fig:fm}}
\end{figure}

The boundary between layer~A and layer~B is defined based on a horizontal averaging scheme
from the stellar convection literature, see, e.g., \citet{nordlund_09}
and \cite{viallet_13}. We define volume-weighted horizontal averages
(denoted by angled brackets) of any quantity $X$ such as velocity,
density, or pressure as follows,
\begin{equation}
\langle X \rangle = \frac{\int X \, \ud \Omega }{\int \ud \Omega}.  
\end{equation}
The quantity $X$ is then decomposed into a mean and a fluctuating component,
\begin{equation}
X = \langle X \rangle + X'.
\end{equation}
For defining the boundary between the convective region and the stable surface region, we consider the turbulent mass flux $f_m$,
\begin{equation} \label{eq:mflux}
f_m = \langle \rho' v_r'\rangle. 
\end{equation}
Inside the convective region, heavier fluid is
advected downwards while fluid that is lighter than average rises
upwards. The turbulent mass flux will, therefore, always be negative
in the convective layer. 
In the overshooting layer outside the PNS convection zone the
situation is reversed, and $f_m$ is positive as
the overshooting, outward-moving plumes, are denser
than their surroundings.  In our calculations we
include the overshooting region in layer~A. To capture
properly both the convective zone and the region of overshooting, we
define the boundary between layer~A and
layer~B as the radius where
\begin{equation} 
 f_m  = \left.  0.1 f_m^\mathrm{max}\right  |_{r>r_\mathrm{max}},
\end{equation} 
where
$f_m^\mathrm{max}$ and $r_\mathrm{max}$ are the maximum value of the turbulent mass flux and the radius at which we find $f_m^\mathrm{max}$, respectively.   
This definition can be more easily understood with the help of a radial profile of $f_m$ as shown in Fig.~\ref{fig:fm} for model s27 at a post-bounce time of 
$192 \, \mathrm{ms}$. Where necessary, we further distinguish between the convective layer
(layer~A1) and the overshooting layer (layer~A2), which are separated
by the radius where $f_m=0$.


\begin{figure*}
\includegraphics[width=0.49\textwidth]{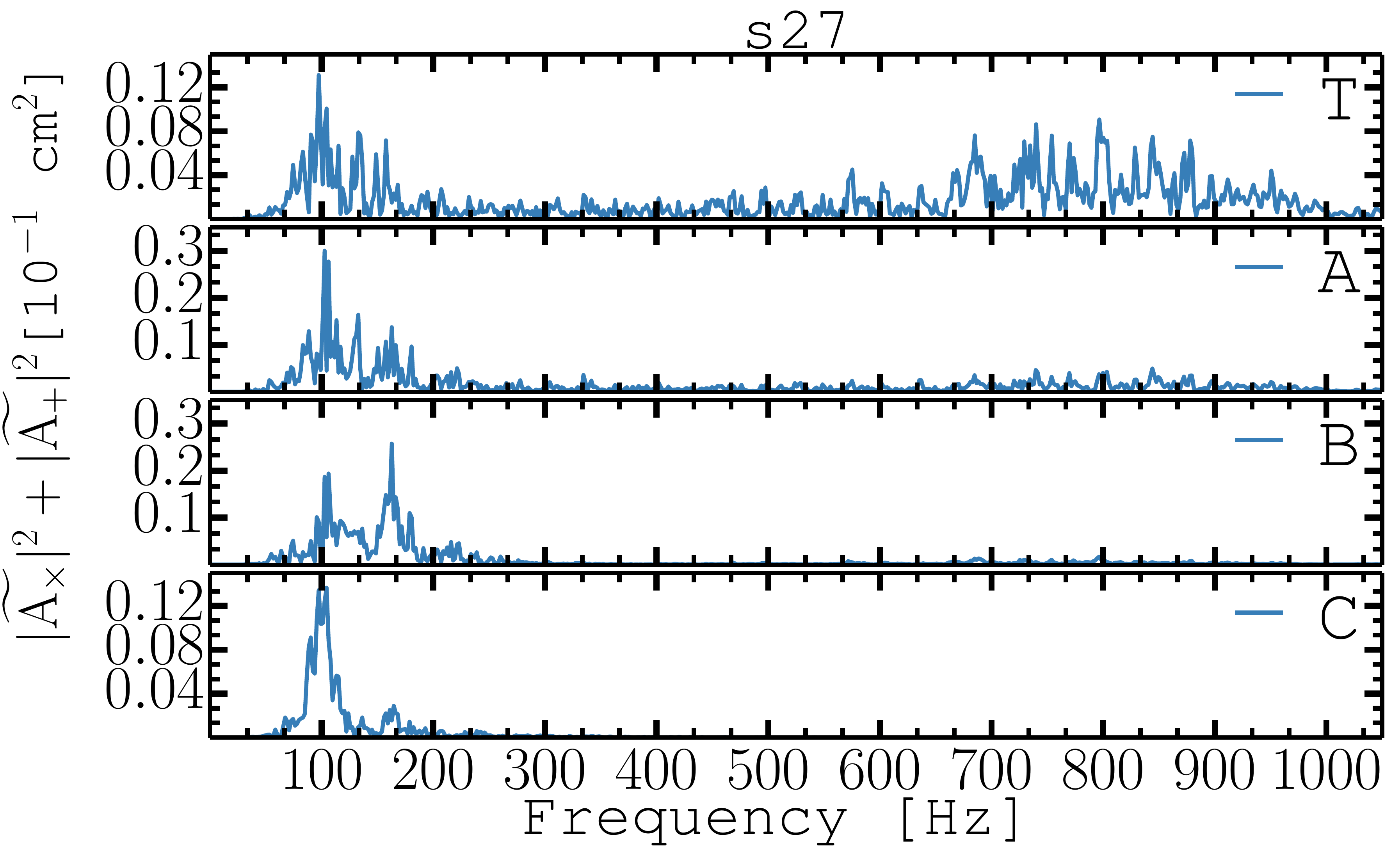}
\includegraphics[width=0.49\textwidth]{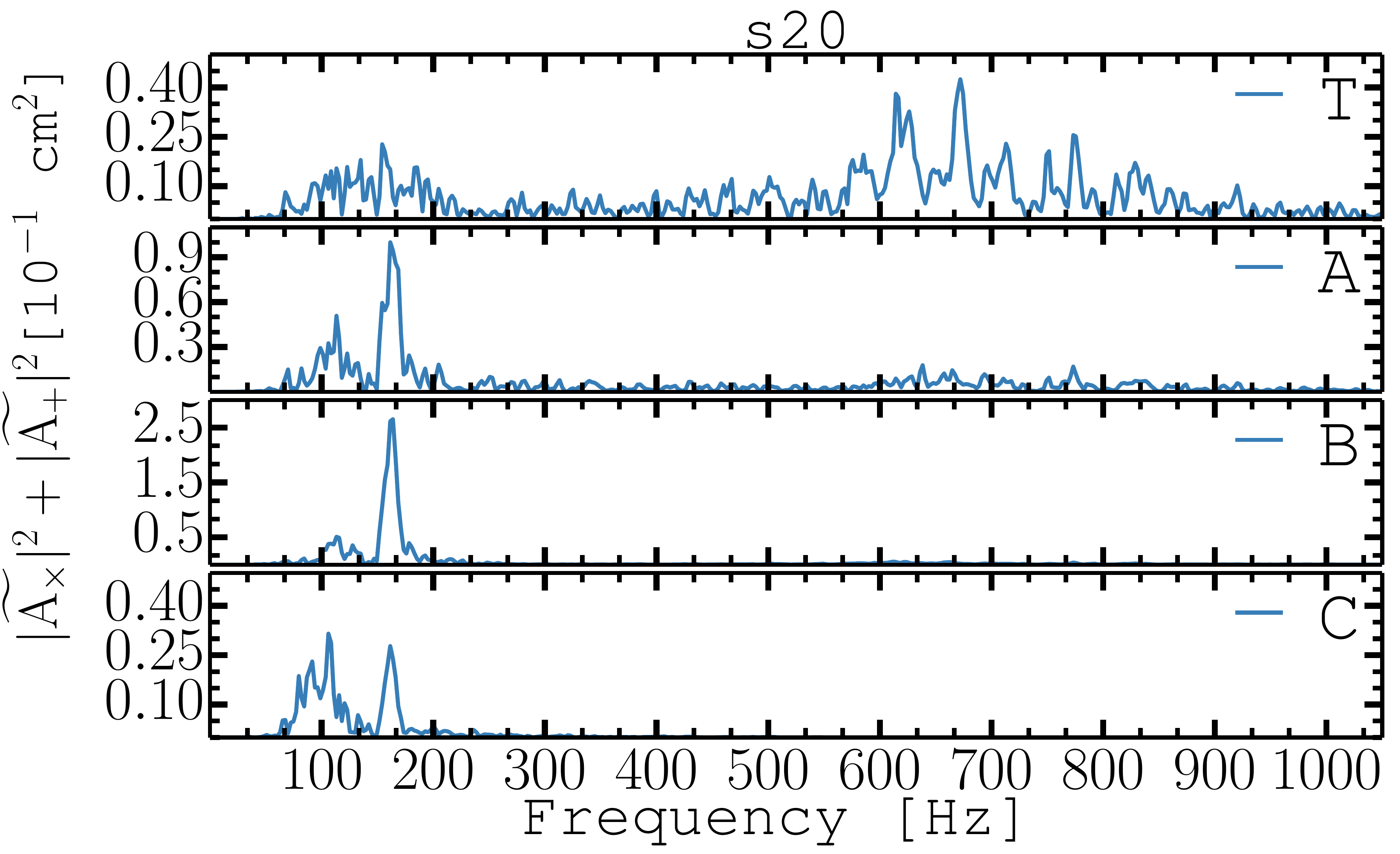}
\\
\includegraphics[width=0.49\textwidth]{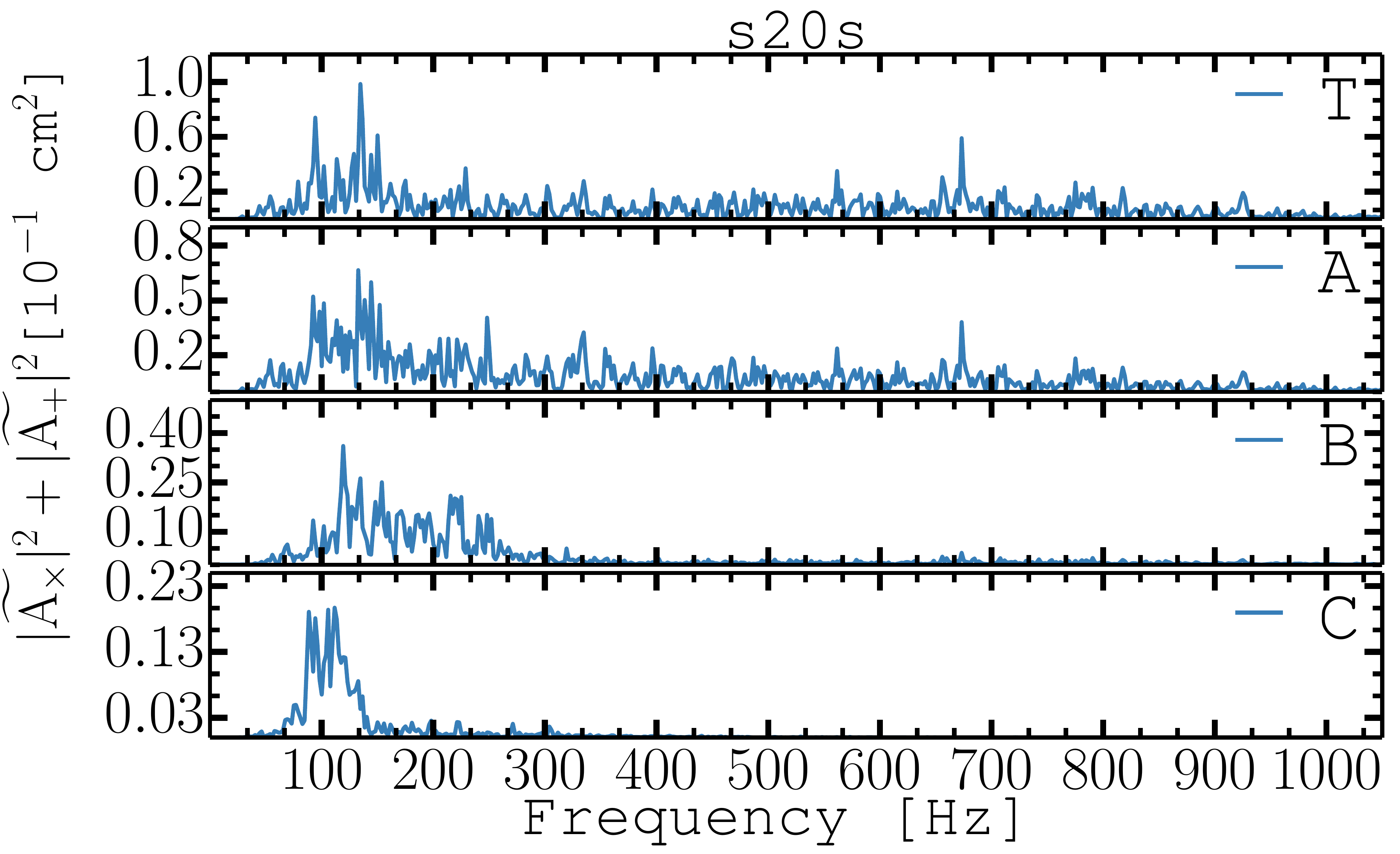}
\includegraphics[width=0.49\textwidth]{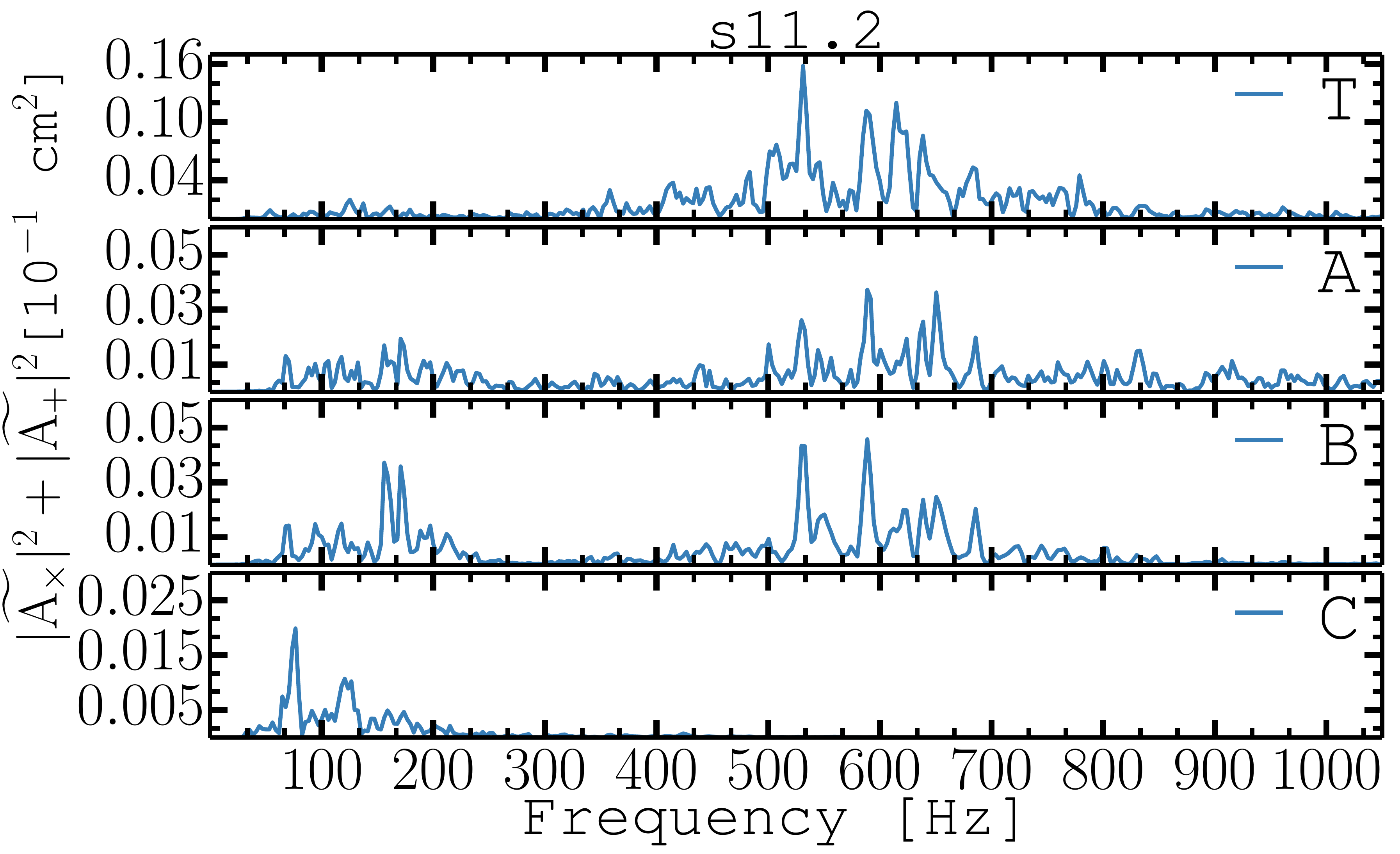}
\caption{Squared Fourier amplitudes of the total volume-integrated GW
  signal and of the signal contributions arising from the three different
  layers of the simulation volume.  
  From the top: Total signal~(T), 
  the PNS convective region and the overshooting layer~(A), the PNS surface layer~(B),
  and the volume between the PNS surface and the outer grid boundary~(C).  
  Top row: the left and right columns show the
  results for models s27 and s20,
  respectively. Bottom row: the left and right columns show the
  results for models s20s and s11.2.
  See Fig.~\ref{fig:PNSski} for a sketch of the three
  regions used for this analysis. The Fourier amplitudes are calculated according to
  Eq.~(\ref{eq:DFT}), Eq.~(\ref{eq:aplus}) and Eq.~(\ref{eq:ax}). 
\label{fig:cuts}}
\end{figure*}

In Fig.~\ref{fig:cuts} we plot the Fourier amplitudes of the GW
amplitudes arising from each individual layer; these amplitudes are
calculated from the full-time signal and for an observer situated at
the pole, corresponding to the left column in
Fig.~\ref{fig:spectrograms}. This figure has to be analysed with some care. Since
we plot the square of the Fourier coefficients one can not add the values of layers A, B, and C 
together and recover the value for the total signal. In addition, artefacts can arise due to
effects at the boundaries between layers, as in the case of model s20 (top right panel of Fig.~\ref{fig:cuts}).
There is an artificially strong peak at 160 Hz, particularly from layer B. We have confirmed that shifting the 
boundary between layers A and B inwards reduces this peak significantly. The exact values of the low-frequency
amplitudes are sensitive to the boundary definition, but the fact that all three layers contribute to
emission below $250 \, \mathrm{Hz}$ is robust. The high-frequency component is less affected by such artefacts since the high-frequency emission is mostly confined to layer A.  

The results of this dissection of the
contributions to the integral in Eq.~(\ref{eq:STFQ}) are somewhat
unexpected.
The high-frequency emission  mostly stems from aspherical
mass motions in layer~A and there is only a
minor contribution from layer~B, which has been posited as
the crucial region for GW emission during the pre-explosion phase in
works based on 2D simulations \citep{marek_08,murphy_09,mueller_13}.
Aspherical mass motions in layer~C hardly contribute
to this component at all.

By contrast, \emph{all three regions} contribute to the low-frequency
signal (i.e.\ emission at frequencies lower than 250 Hz) to a similar degree. This is also surprising
if the dominant frequency of this component appears to be set by the SASI
as speculated before. In this case, one might expect that
the fluid motions responsible for GW emission are propagating
waves in layer~C and perhaps layer~B,
where the conversion of vorticity perturbations into acoustic
perturbations occurs in the SASI feedback cycle. 

\subsection{Origin of High-Frequency Emission}
What do these findings imply about the physical mechanisms that give
rise to GW emission and determine their frequency? Let us first
address the high-frequency signal. Recent 2D studies have connected GW
emission at $\mathord{\gtrsim} 500 \, \mathrm{Hz}$ to oscillatory modes
(g-modes) excited either in the PNS surface (layer B) from above by
downflows impinging onto the PNS
\citep{marek_08,murphy_09,mueller_13}, or from below by PNS
convection \citep{marek_08,mueller_e_12,mueller_13}. Prior to shock
revival, the excitation of oscillations by mass motions in the gain layer was found
to be dominant, with PNS convection taking over as the dominant
excitation mechanism only after the onset of the explosion \citep{mueller_e_12,mueller_13}.
The typical angular frequency of such
processes is roughly given by the Brunt-V\"{a}is\"{a}l\"{a} frequency, $N$,
in the convectively stable region between the gain region and the PNS convection zone,
\begin{equation} \label{eq:BV}
N^2 = \frac{1}{\rho} \frac{\partial \Phi}{\partial r} \left [ \frac{1}{c_s^2} \frac{\partial P}{\partial r} - \frac{\partial \rho}{\partial r} \right ],
\end{equation}
where $c_s$ is the sound speed. \citet{mueller_13} further investigated the dependence
of this frequency on the mass $M$, the radius $R$, and the surface temperature $T$ of the PNS to explain the secular
increase of $N$ during the contraction of the PNS and
a tendency towards higher frequencies for more massive neutron stars.

\begin{figure}
\includegraphics[width=0.45\textwidth]{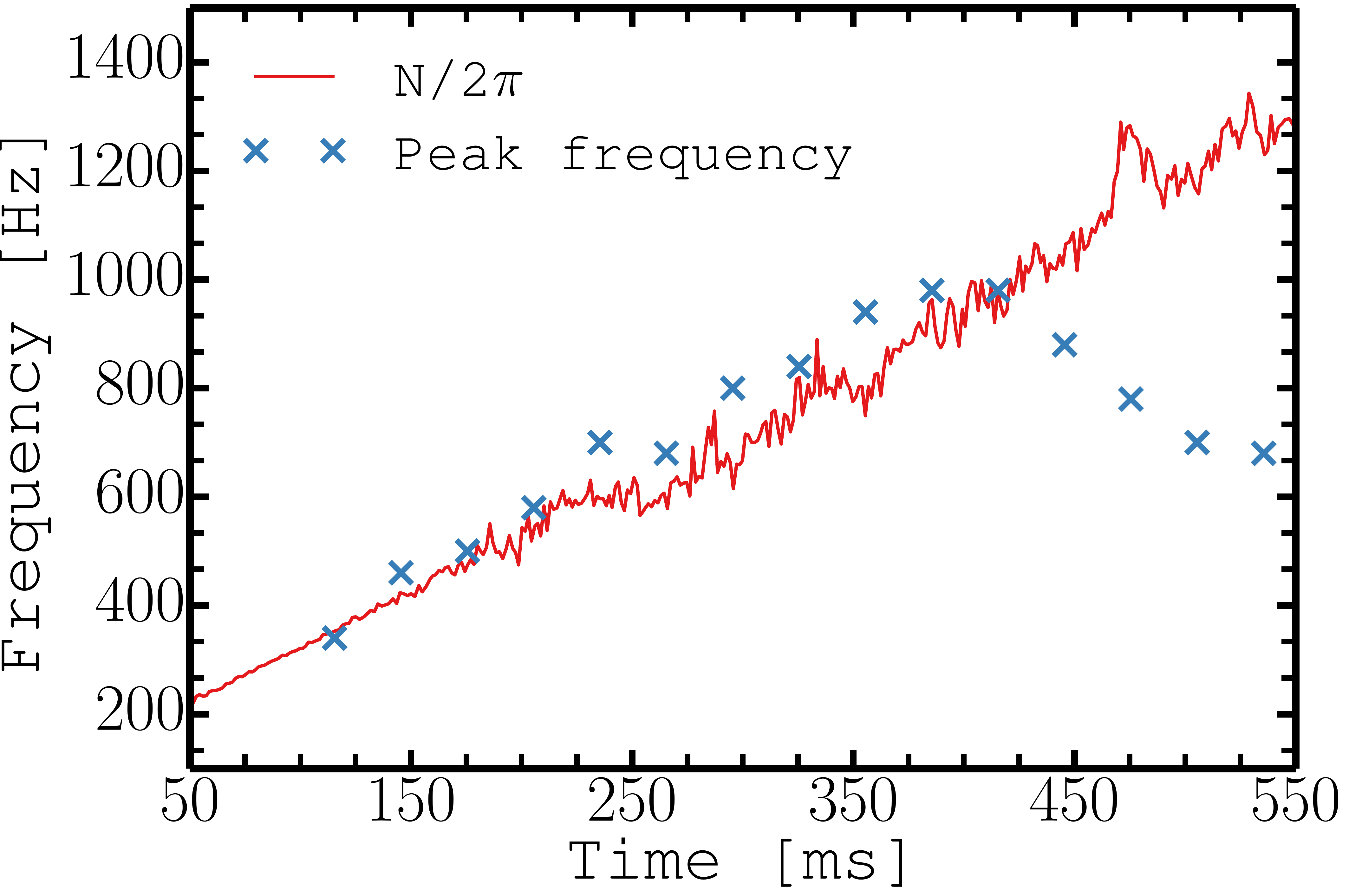}
\caption{Frequency of strongest GW emission
above $250 \, \mathrm{Hz}$ in the spectrogram of model s27 as a function of time (blue crosses).
We also plot the expected characteristic frequency of GW emission excited by buoyancy effects
in the PNS surface layer~B and the overshooting layer~A2 (red curve, see definition in Eq.~(\ref{eq:BV}).
The exact value of $N$ depends on the radius where Eq.~(\ref{eq:BV}) is evaluated, 
but we find similar numerical values within layers~B and A2. Note that the trends seen for model s27 are common to all our models.
\label{fig:fpeak}}
\end{figure}

Our results confirm that the peak frequency of the high-frequency GW emission
is still set by the Brunt-V\"{a}is\"{a}l\"{a} frequency in 3D and therefore point to a
similar role of buoyancy forces in determining the spectral structure
of the high-frequency component. As shown in Fig.~\ref{fig:fpeak} for
model s27, we find very good agreement between the peak GW
frequency, $f_\mathrm{peak}$, and the Brunt-V\"{a}is\"{a}l\"{a}
frequency, $N$, calculated at the outer boundary of the overshooting
layer (the boundary between layers~A and B).
Here, $f_\mathrm{peak}$ denotes the frequency with the highest Fourier amplitude above $250 \, \mathrm{Hz}$.
Superficially, there appears to be a discrepancy at post-bounce times
later than $400 \, \mathrm{ms}$, where $f_\mathrm{peak}$ seems to decrease
again. This, however, is purely an artefact of the
sampling rate of $0.5 \, \mathrm{ms}$ in the simulations, which results
in a Nyquist frequency of $1000 \, \mathrm{Hz}$. The peak frequency is
therefore aliased into the region below $1000 \, \mathrm{Hz}$. If
this is taken into account, there is in fact good agreement between
the Brunt-V\"{a}is\"{a}l\"{a} frequency of $\mathord{\sim} 1300
\, \mathrm{Hz}$ and the aliased peak GW frequency of 
$\mathord{\sim} 700\, \mathrm{Hz}$ at the end of the simulation.

The dominant excitation mechanism for these oscillatory motions in layers A and B is, however,
remarkably different from previous 2D models.
While allowing for a minor contribution from PNS
convection to the total signal during the pre-explosion phase, most of the signal in 2D is
found to originate from oscillations in layer~B that are excited by convective plumes and/or the
downflows of the SASI \citep{marek_08,murphy_09,mueller_13}.
In this case, one would expect that the excited oscillation modes have
large amplitudes mostly in the surface layer and that this layer
contributes significantly to the GW signal. This is not the case in
3D as shown by Fig.~\ref{fig:cuts}.
The dominant contribution from layer~A rather suggests that
oscillatory modes are predominantly excited from below by aspherical mass
motions in the PNS convection zone and are confined mostly to the overshooting layer, which
acts as frequency stabiliser.
{\comment This assessment is also compatible with the temporal
  evolution of the amplitudes and the power in the spectrograms.
  The amplitudes (Fig.~\ref{fig:amps}) show modest
  temporal variations after an initial GW-quiet
  phase of $150 \text{-}200 \, \mathrm{ms}$ and little response
  to strong activity of SASI and convection, which argues
  against efficient excitation of surface g-modes by motions
  in the gain region. The spectrograms
  (Fig.~\ref{fig:spectrograms}) point to the
  same conclusion, e.g.\ high-frequency emission is practically
  absent during the first phase of SASI activity in  s27.}

To confirm the crucial role of layer~A2, we excluded
this region from our analysis and found a large reduction of the
energy carried away by high-frequency GWs,
\begin{equation}
E_{GW} \sim \int_{250 \, \mathrm{Hz}}^{1100\, \mathrm{Hz}} \frac{\ud E}{\ud f} \, \ud f.
\end{equation}
For model s27, we find a reduction of the GW energy
by roughly a factor of two when excluding the overshooting layer~A2. 
It is
remarkable that the deeper regions of the PNS convection zone (layer~A1)
nonetheless contribute to the high-frequency signal with similar
frequencies: There is no apparent reason for a correlation between the
convective overturn time $T_\mathrm{conv}$ (which sets the natural
frequency for GWs from the bulk of the PNS convection zone as
$1/T_\mathrm{conv}$) and the Brunt-V\"{a}is\"{a}l\"{a} frequency in
the overlying stable region.

\subsection{\comment Comparison of High-Frequency Emission in 2D and 3D}
\begin{figure*}
\includegraphics[width=0.48\textwidth]{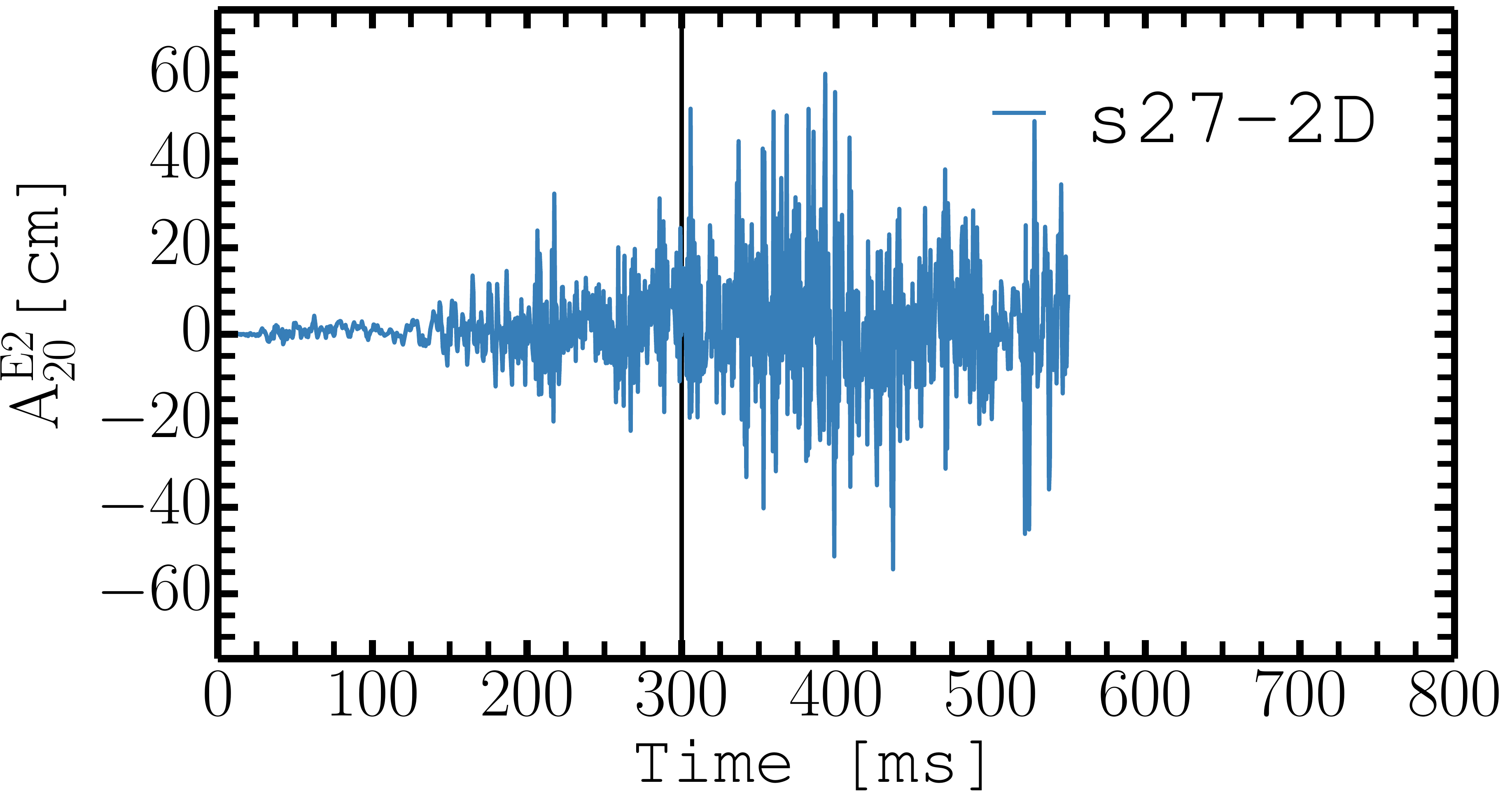}
\includegraphics[width=0.48\textwidth]{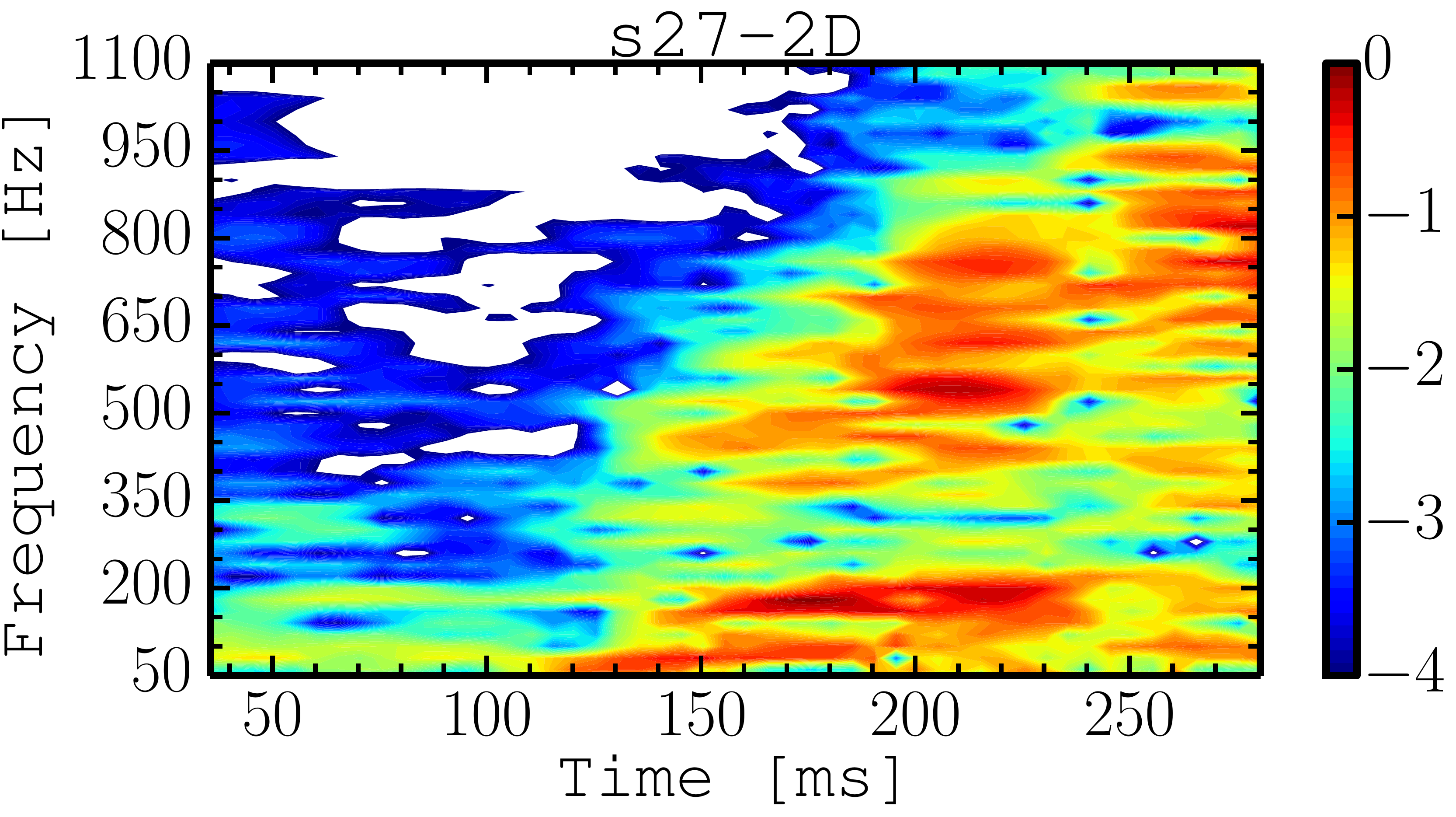} \\
\includegraphics[width=0.48\textwidth]{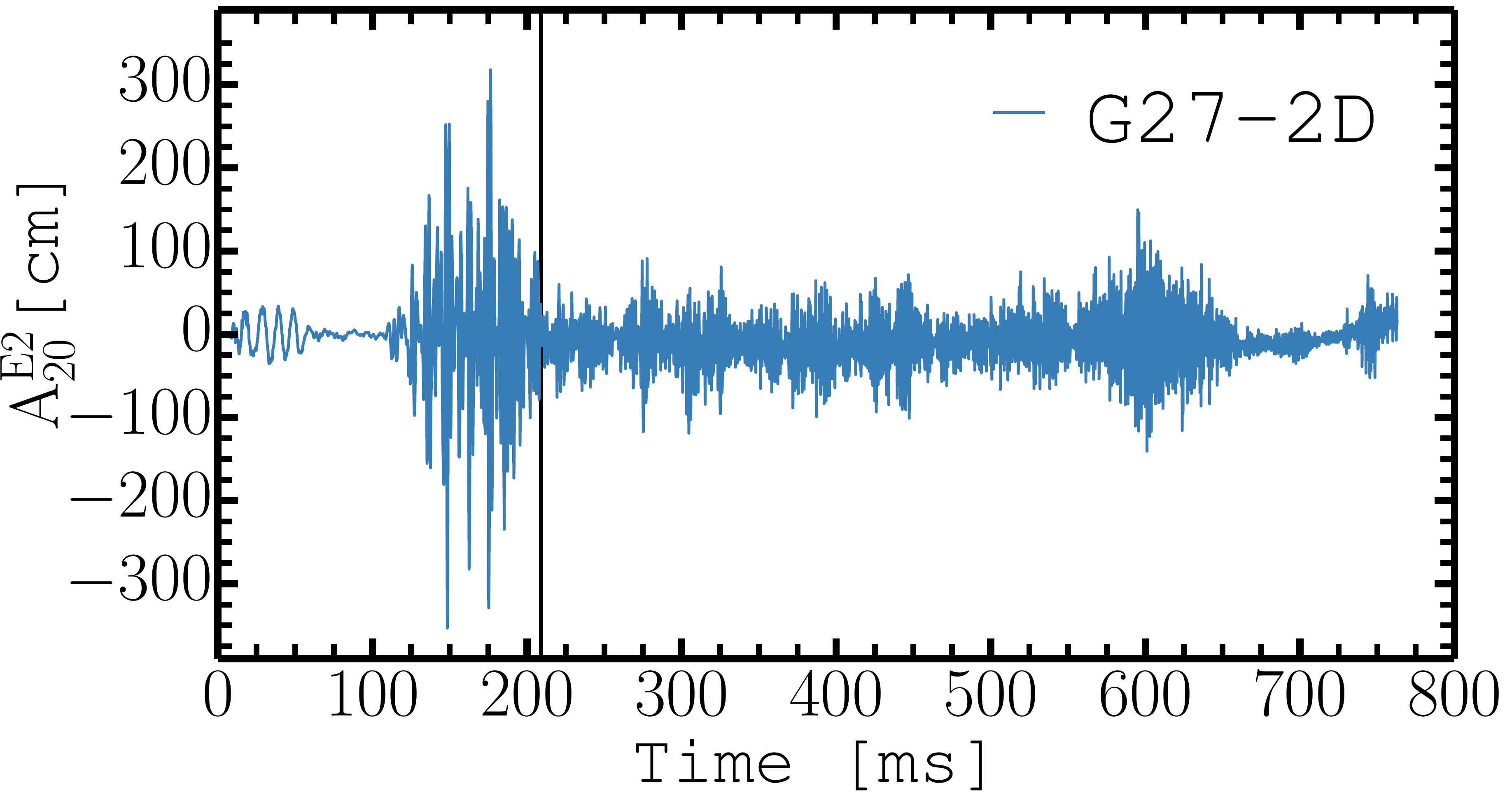}
\includegraphics[width=0.48\textwidth]{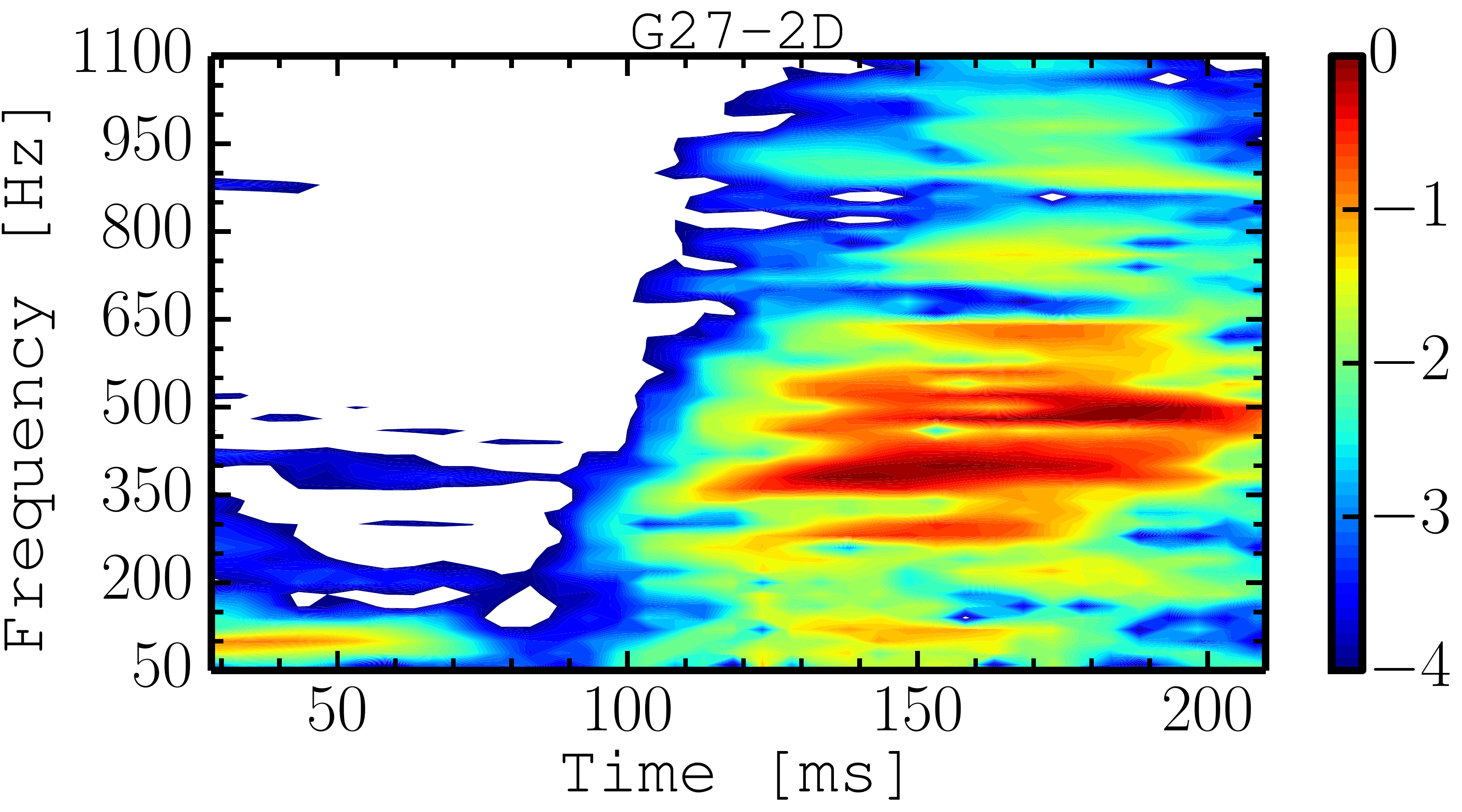}
\caption{The GW amplitude, $\mathrm{A}_{\mathrm\mathrm{E2}}^{\mathrm{20}}$, as a function of time after bounce (left)
and amplitude spectrograms in in logarithmic scale (right) for the two 2D models s27-2D (top row) and G27-2D (bottom row). 
For a useful comparison with the corresponding non-exploding 3D model, we only
show the spectrograms for the time between bounce and the onset of the explosion.
\label{fig:s272d}}
\end{figure*}

\begin{figure*}
\includegraphics[width=0.45\textwidth]{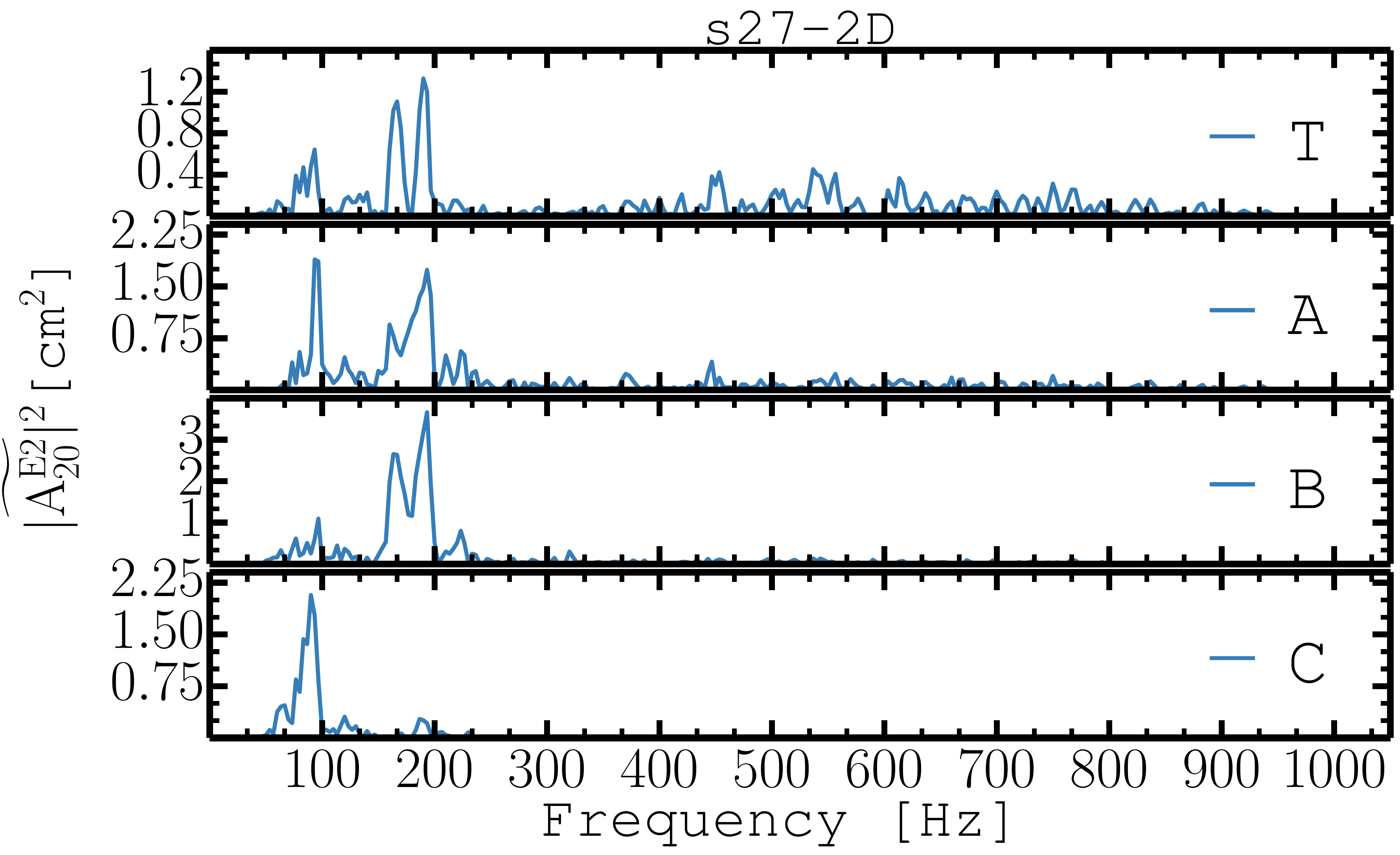}
\includegraphics[width=0.45\textwidth]{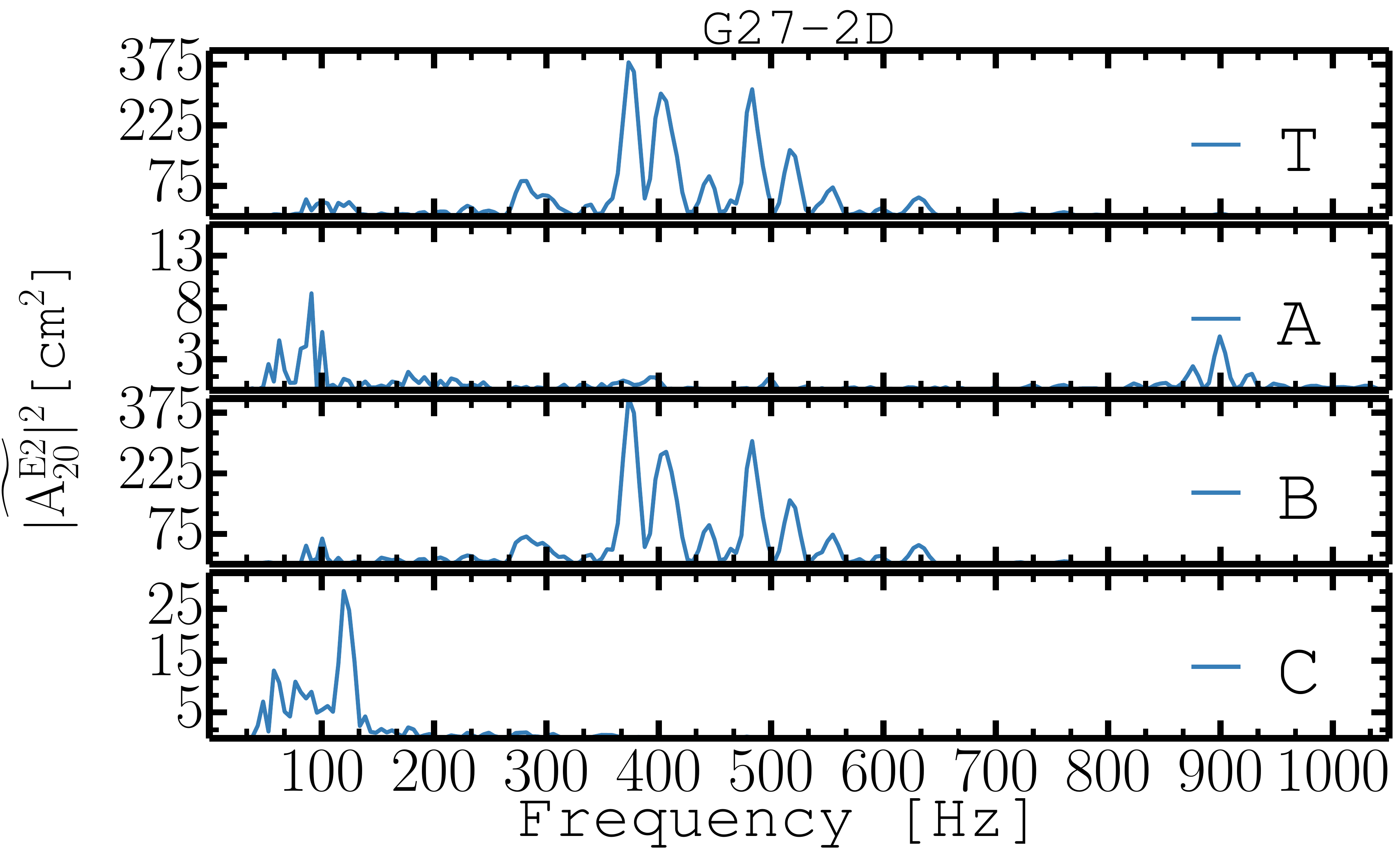}
\caption{Squared Fourier amplitudes of the total volume-integrated GW signal
  and of the signal arising from the three different layers of the simulation volume,
  for the 2D model G27-2D of \citet{mueller_13} (right) and s27-2D (left). 
  From the top: total signal (T), layer A, layer B, and layer C. 
  The Fourier amplitudes, $\widetilde{\mathrm{A}_{\mathrm\mathrm{E2}}^{\mathrm{20}}}$, are calculated based on Eq.~(\ref{eq:DFT}) and Eq.~(\ref{eq:2dquad}).
\label{fig:2dcut}}
\end{figure*}

To further illustrate the differences between previously published 2D waveforms and our 3D results, we
re-analyse the GW signal from model G27-2D of \citet{mueller_13} using
the STFT and the decomposition of the computational volume into three
different regions. G27-2D is a 2D model based on the same $27 M_\odot$
progenitor used in our simulations, with the same equation of state
and the same neutrino treatment, the only major difference being the
treatment of GR: In G27-2D, the equations of radiation hydrodynamics in
the ray-by-ray-plus approximation are solved in their general
relativistic formulation assuming a conformally flat metric, whereas
the pseudo-Newtonian approach of \citet{marek_06} was used for model s27
in 3D. The bottom row of Fig.~\ref{fig:s272d} shows the GW signal and amplitude spectrogram for model G27-2D.
Fourier amplitudes for the signal from the three regions are shown in right panel of Fig.~\ref{fig:2dcut}
{\comment for the period up to the onset of the explosion $210 \, \mathrm{ms}$ after bounce}.

Comparing the spectrograms of model G27-2D with those of the 3D models
(Figs.~\ref{fig:spectrograms} and \ref{fig:s272d}), we find that the
``quiet zone'' between the high-frequency and low-frequency components
of the signal is not {\comment visible} in model G27-2D, which is in
agreement with the wavelet analysis of \citet{mueller_13} who also
found a more broad-banded signal during phases of strong SASI and
convection, with the Brunt-V\"{a}is\"{a}l\"{a} frequency providing
more of an upper limit rather than a sharply defined peak frequency
during such phases. In model G27-2D, there is strong emission between
$250 \, \mathrm{Hz}$ and $750 \, \mathrm{Hz}$, and the relative
contribution of the low-frequency signal below $250 \, \mathrm{Hz}$ is
smaller (which will be discussed in more detail in
Section~\ref{sec:lowfreq}).  The decomposition of the integration
volume into three layers (Left panel of Fig.~\ref{fig:2dcut}) reveals
that the main contribution to the high-frequency signal stems from the
PNS surface (layer~B).  In addition to the broadband emission between
$250 \, \mathrm{Hz}$ and $750 \, \mathrm{Hz}$, there are also two
narrow emission peaks centred around $800 \, \mathrm{Hz}$ and $900 \,
\mathrm{Hz}$.  This emission is the result of oscillations deep in the
PNS core that are excited by PNS convection.

\begin{figure}
\includegraphics[width=0.49\textwidth]{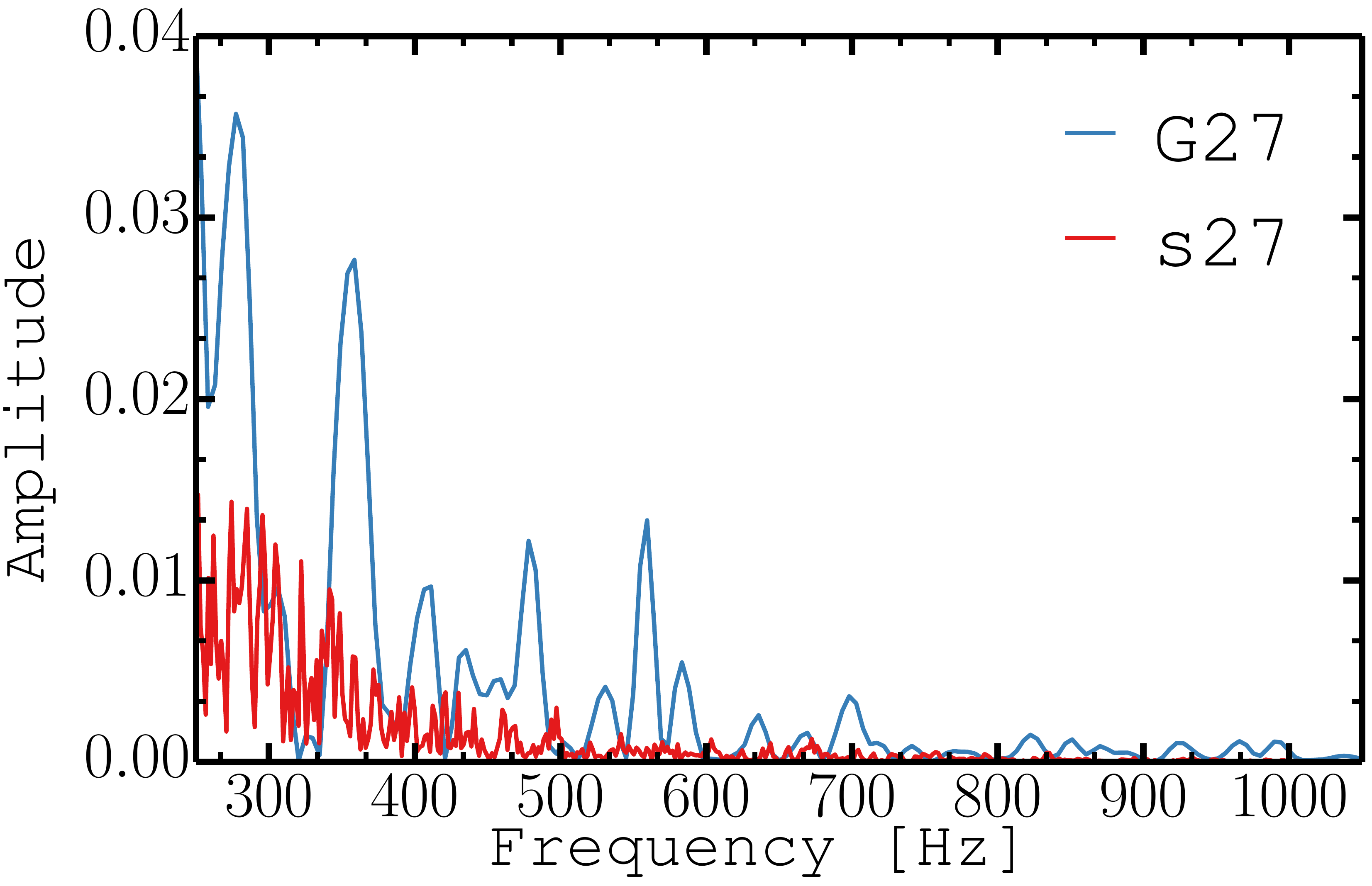}
\caption{Normalised Fourier amplitudes of the GW signal from layer~C
for the 2D model G27-2D and the 3D model s27. Each curve is normalised by its respective maximum to account for
the difference in magnitude between 2D and 3D. Note that the maxima lie outside of the frequency domain shown in this figure.
\label{fig:post3v2}}
\end{figure}

{\comment Different from model G27-2D and other recent 2D models found in
  the literature \citep{marek_08,murphy_09,mueller_13}, a 2D version of
  model s27 (model s27-2D, simulated with \textsc{prometheus-vertex})
  more closely resembles the 3D models presented in our study during
  the pre-explosion phase (i.e.\ up to $300\, \mathrm{ms}$ after
  bounce) in terms of its spectrogram (top right panel in
  Fig.~\ref{fig:s272d}) and time-integrated spectrum (left panel of
  Fig.~\ref{fig:2dcut}).  The spectrogram of the model (top right
  panel of Fig.~\ref{fig:s272d}) shows the same two signal components
  that we found in our 3D models. There is a high-frequency and a
  low-frequency component and a frequency band separating the two
  components where the emission is much weaker (between 250 and $350\,
  \mathrm{Hz}$).  The \emph{relative} contributions to the total
  signal from layers A, B, and C are roughly the same in models s27 and
  s27-2D (see left panel of Fig.~\ref{fig:2dcut} and top left panel of
  Fig.~\ref{fig:cuts}).  Judging from the time-integrated signals from
  the pre-explosion phase, the only noteworthy difference between
  models s27 and s27-2D appears to consist in an overall reduction of
  the amplitudes by about a factor of ten (or by a factor of 100 in
  squared amplitudes as shown in Figs.~\ref{fig:cuts} and
  \ref{fig:2dcut}) in 3D (s27) compared to s27-2D in all regions
  across the entire spectrum. 

  The fact that the signal from layer~B is not very strong
  in s27-2D makes it difficult to determine the impact of 2D effects
  on mode excitations by motions in the gain
  region as opposed to motions in the PNS convection zone. The
  modes excited by motions in the gain region need to be very strong for
  the emission from layer~B to clearly stick out in time-integrated
  spectra as in G27-2D. The spectrograms, however, show that
  the excitation of surface g-modes by the SASI is more efficient
  in s27-2D than in the 3D model s27: In s27, significant emission in
  the high-frequency band is absent up to $\mathord{\sim} 210\, \mathrm{ms}$
  after bounce despite strong SASI activity (which shows up
  in the low-frequency band), whereas s27-2D shows noteworthy
  emission in the high-frequency band during this phase.
In 3D, we thus find i) a suppression of the signal originating
from PNS convection (layer~A) by a factor of $\mathord{\sim 10}$,
and ii) an even more efficient suppression of
any high-frequency GW emission due to mode excitation
by the SASI in the spectrograms. 

There are presumably several reasons why the excitation of
oscillations in the PNS surface layer is found to be more efficient
in 2D than in 3D.}  First, the
inverse turbulent cascade \citep{kraichnan_76} and the suppression of
the Kelvin-Helmholtz instability at the edge of supersonic downflows
(see \citealt{mueller_15b} and references therein) lead to an
artificial accumulation of the turbulent energy on large scales in 2D
supernova simulations \citep{hanke_12,abdikamalov_15} and higher
impact velocities of the downflows
\citep{melson_15a,mueller_15b}.  Thus, both the amplitudes as well as
the mode overlap of the forcing with the excited $l=2$ oscillation
modes are higher in 2D.  However, while the
excitation of g-modes in the surface layer is strongly suppressed in
3D, there is still some residual g-mode activity
\citep{melson_15a,mueller_15b}.  {\comment Furthermore,} we must also
consider the frequency structure of the forcing.
Fig.~\ref{fig:post3v2} shows considerable high-frequency emission from
layer~C in 2D, which is indicative of violent large-scale (i.e.\ with
an $l=2$ component) mass motions on time-scales considerably
\emph{shorter} than the SASI period or the convective overturn
time-scale. The lack of such high-frequency GW activity from layer~C
in 3D indicates that the downflows in 3D are not as strongly distorted
by intermediate-scale eddies and that they vary less on short
time-scales.  Fig.~4 of \citet{melson_15a}, which shows 2D and 3D
simulations of a successfully exploding $9.6 M_\odot$ model, further
illustrates this difference between the frequency structure of the
post-shock flow in 2D and 3D: In 3D the angle-averaged radial velocity
profiles of the infalling material appear smooth. On the other hand,
intermediate-scale eddies with fast time variations are clearly
visible in 2D.  With a typical time scale on the order of $t \sim
1\ldots 10 \, \mathrm{ms}$, corresponding to a frequency range of $f
\sim 100 \ldots \,1000 \, \mathrm{Hz}$, the eddies in 2D cause a more
``impulsive'' forcing with a broad frequency spectrum.  The different
frequency structure of the forcing in 2D and 3D is then reflected in
the excited PNS surface oscillations: In 2D, where the frequency
spectrum of the forcing overlaps with the natural g-mode frequency, we
see \emph{resonant} excitation of free (high-frequency) g-mode
oscillations.  This is not the case in 3D, but we still see strong
\emph{non-resonant} excitation of forced g-modes at low frequencies in
the PNS surface (see Section~\ref{sec:lowfreq}). 

{\comment It is noteworthy that some of these aforementioned effects
(e.g.\ the different dynamics of supersonic downflows) depend on the Mach number
of the flow and are therefore
only relevant for the damping of the excitation of surface g-modes by motions from
the gain region. By contrast, Mach-number dependent effects
like the inhibition of the Kelvin-Helmholtz instability in 2D
will not affect the excitation of oscillation modes by PNS convection,
which is strongly subsonic. This explains why 3D effects
strongly quench g-mode excitation from SASI and
hot-bubble convection, but lead to a more
modest reduction of GW emission due to PNS convection, which
therefore becomes the dominant source of high-frequency GW emission in
our 3D models.
}

{\comment Without a 3D model corresponding to G27-2D, it is not yet
  possible to decide whether the suppression of surface g-mode
  excitation is \emph{generically} strong enough to make it a
  subdominant source of GW emission. The critical feature in the
  post-shock flow that is responsible for strong GW emission from
  layer~B in G27-2D is the development of very strong non-linear SASI
  activity with the presence of stable supersonic downflows, which
  reach Mach numbers of $1.5$ already $110 \, \mathrm{ms}$ after
  bounce. Since the coupling of perturbations in the gain region to
  the surface g-modes is more effective for high Mach number flow
  \citep[cp.][]{goldreich_90,lecoanet_12}, and since the kinetic
  energy in non-radial motions reaches large values of up to $6 \times
  10^{49} \, \mathrm{erg}$ (compared to pre-explosion values of
  $\mathord{\lesssim 1.5} \times 10^{49} \, \mathrm{erg}$ in s27
and s27-2D,
  cp.\ \citealp{hanke_13}), the resulting GW amplitudes from surface
  g-modes in G27-2D are one order of magnitude larger  than in s27-2D
and two orders of magnituder larger than in s27, which only
  develops mildly non-linear SASI activity during the pre-explosion
  phase. The GW emission from layer~B thus clearly dominates the
  time-integrated spectrum of G27-2D.

 Under the conditions that obtain in model G27-2D (fully
  developed non-linear SASI/convection with high Mach numbers), 2D/3D
  differences in the behaviour of the downflows due to the inhibition
  of the Kelvin-Helmholtz instability in 2D tend to become more
  pronounced, which suggests that very little excitation of surface
  g-mode by the SASI may survive in 3D even for a model with similar
  dynamics to G27-2D.  This will need to be confirmed by future 3D
  simulations, however.  }

\begin{figure*}
\includegraphics[width=0.99\textwidth]{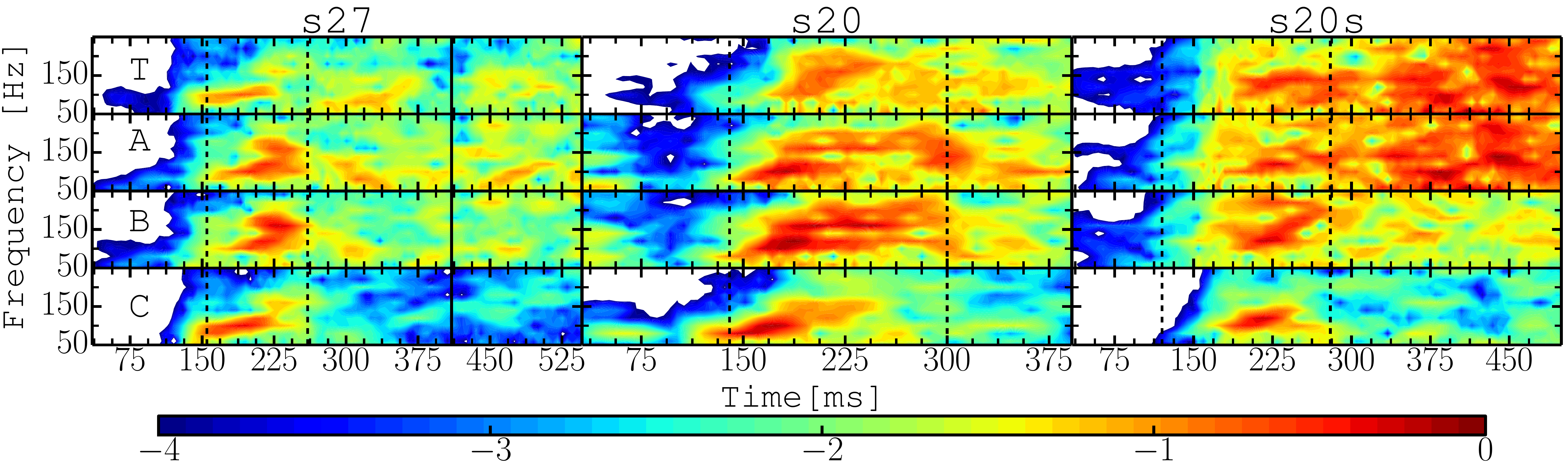}
\caption{Amplitude spectrograms, for a sliding window of 50 ms, of the low-frequency GW signal arising from the three different layers,
  summed over the two polarisation modes ($|\widetilde{A}_+|^2 + |\widetilde{A}_\times|^2$). 
  From the top: total signal (T), layers~A, B, and C. Columns
  are ordered by progenitor (left: s27, middle: s20, right: s20s). 
  See Fig.~\ref{fig:PNSski} for a sketch of the three
  regions. The observer is chosen to be the north pole in the computational grid.
  As in Fig.~\ref{fig:amps} and Fig.~\ref{fig:spectrograms}, strong SASI episodes are bracketed by vertical lines. 
  The colour scale is logarithmic and all panels have been normalised by the same global factor that has also been used for Fig.~\ref{fig:spectrograms}.
\label{fig:lowspec}
}
\end{figure*}

\begin{figure}
\includegraphics[width=0.48\textwidth]{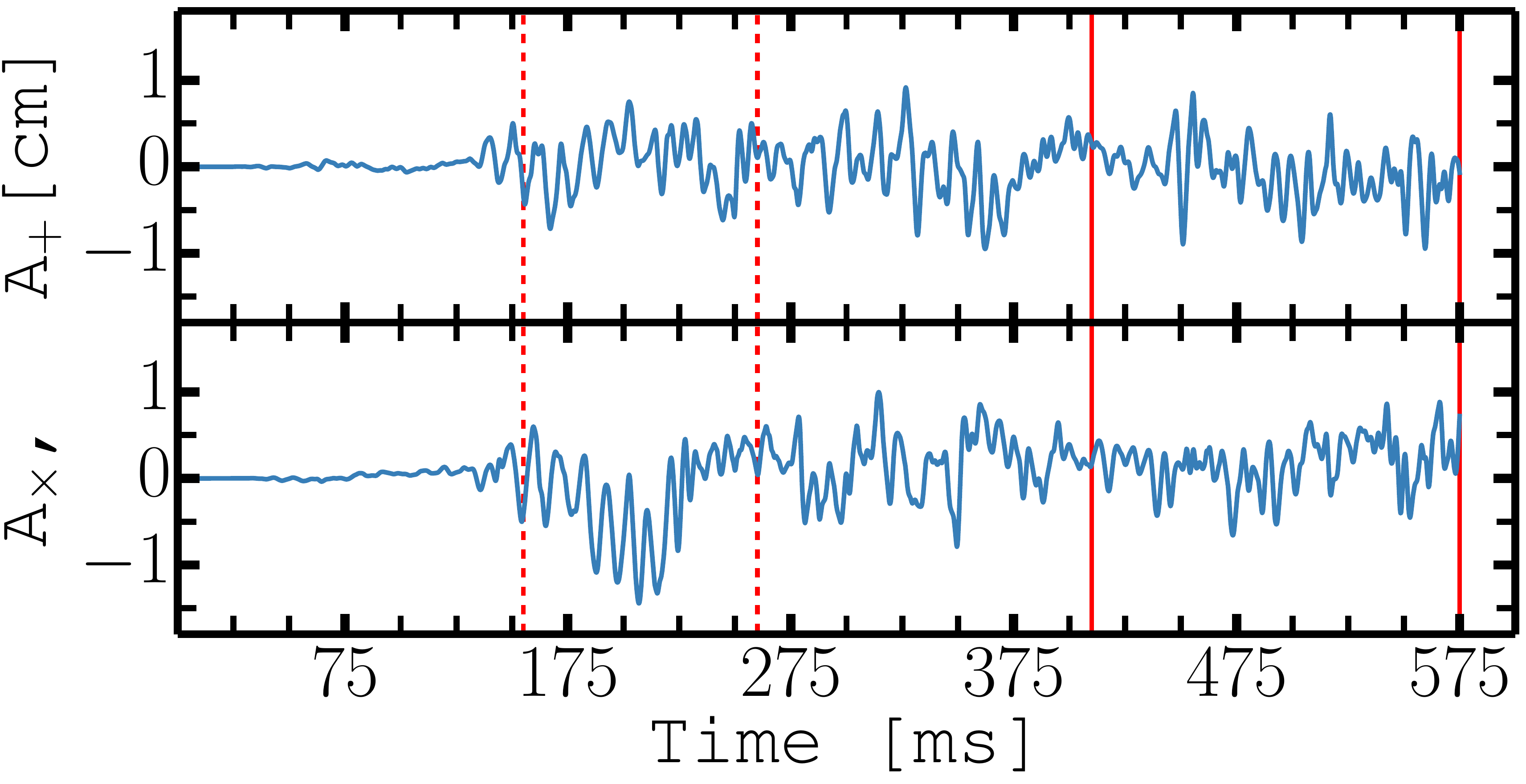}
\caption{The GW amplitudes $A_+$ and $A_{\times}$ of model s27, for an observer situated along the z-axis of the computational grid,
after a low-pass filter has been applied. The time axis indicates time since core bounce.
\label{fig:lowpass}}
\end{figure}

\begin{figure}
\includegraphics[width=0.49\textwidth]{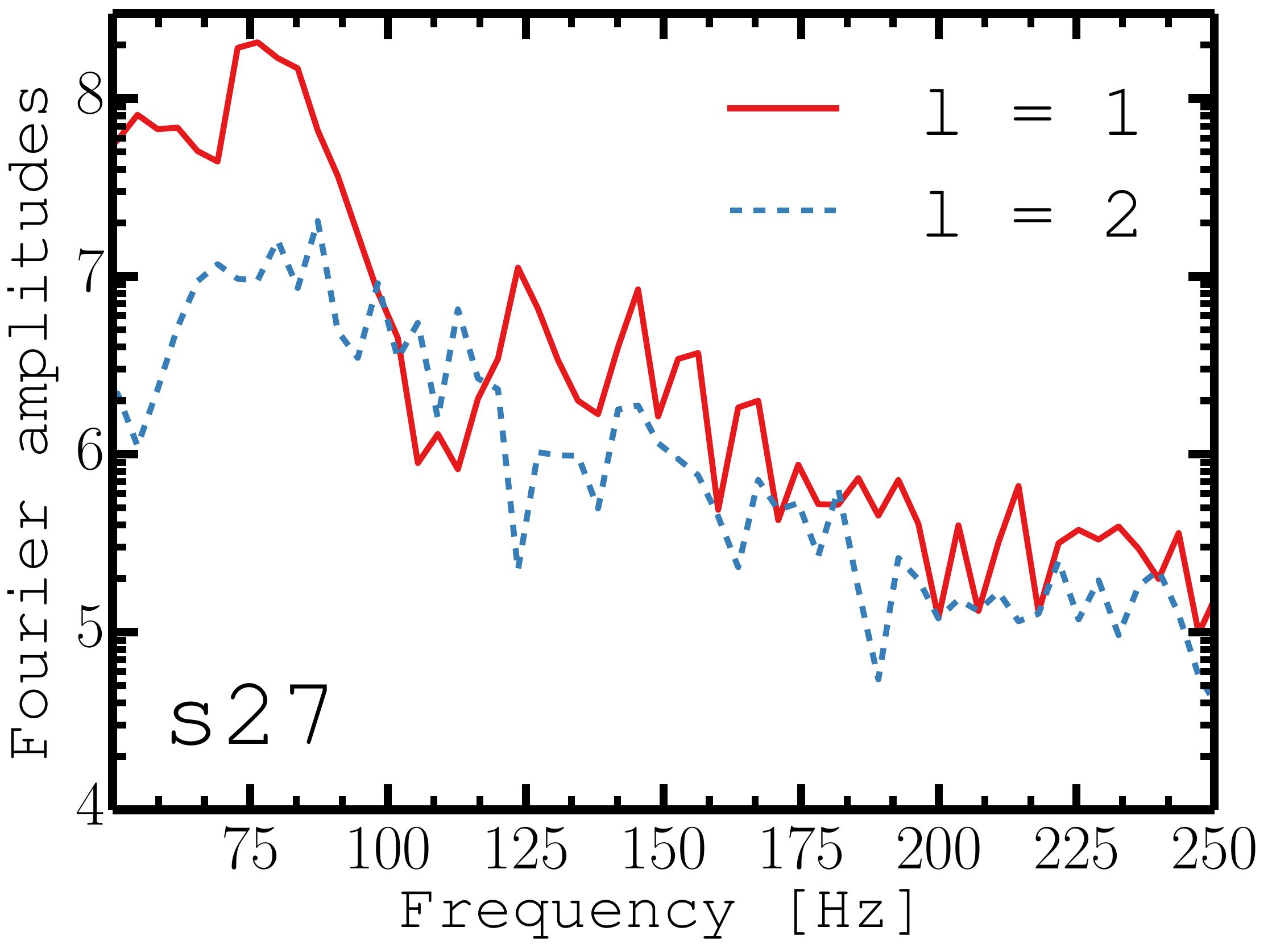}
\\
\includegraphics[width=0.49\textwidth]{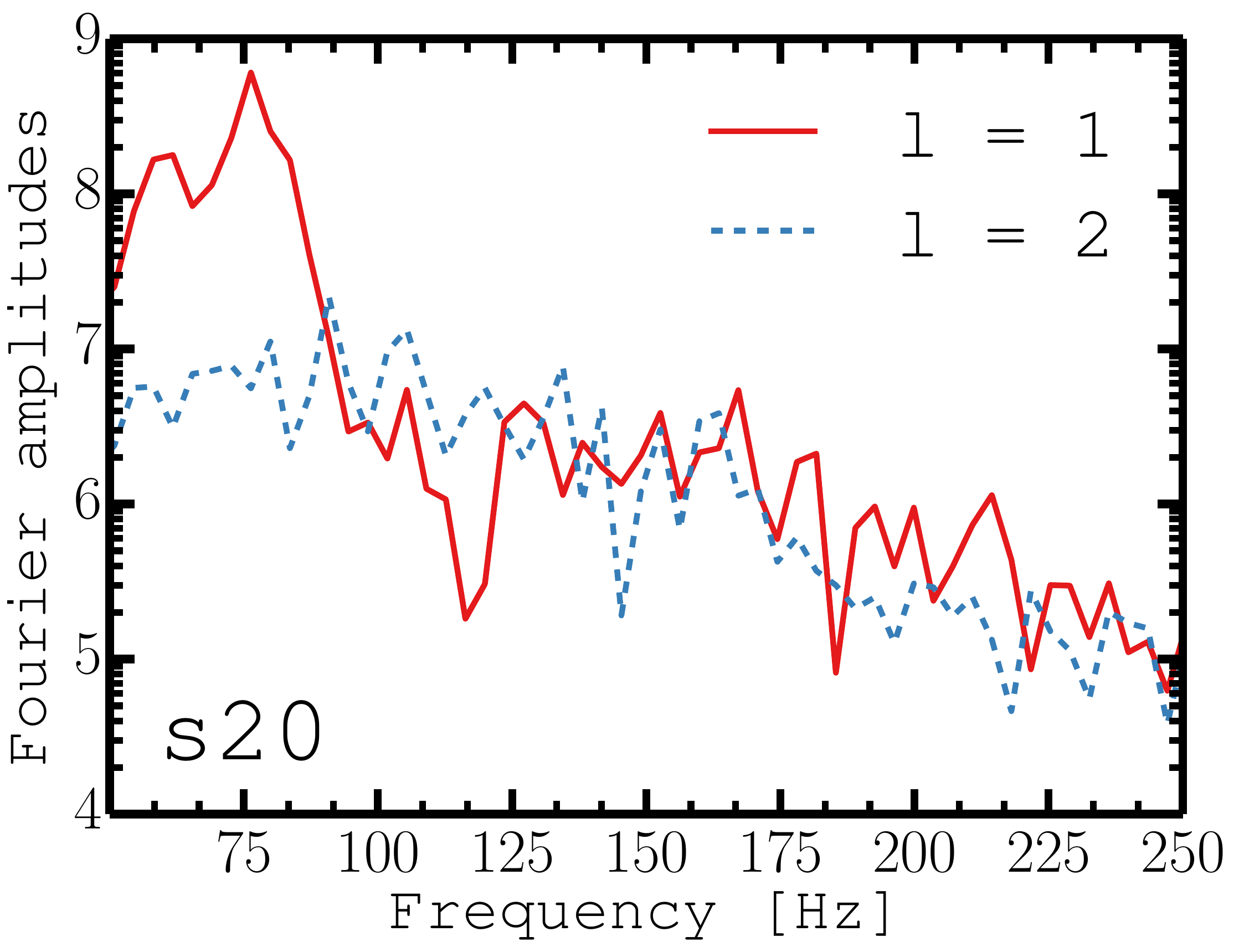}
\caption{Squared Fourier amplitudes, in logarithmic scale, for the $l=1$ and $l=2$ components of the
  expansion of the shock position into spherical harmonics.  The
  Fourier amplitudes have been calculated for the time window between
  $100 \, \mathrm{ms}$ and $350 \, \mathrm{ms}$ after core bounce. 
  The upper panel shows model s27 and the bottom panel the results for model s20.
  The curves in both panels have been normalised by the same factor.
\label{fig:sfreq}}
\end{figure}
  
\subsection{Origin of the Low-Frequency Signal}
\label{sec:lowfreq}
The strong low-frequency signal seen in the more massive models (s20, s20s, s27)
is apparently closely connected to SASI activity.
Note that the convection-dominated s11.2 model show
some stochastic low-amplitude GW emission at low frequencies (see
Fig.~\ref{fig:spectrograms}), which is, however, much less pronounced
compared to the high-frequency component.
To address the origin of
the low-frequency component, we show spectrograms of the GW signal below
$250 \, \mathrm{Hz}$ from each of the three analysis regions for models
s20, s20s, and s27 in Fig.~\ref{fig:lowspec}.

The apparent temporal correlation of the low-frequency emission
with the SASI suggests the following plausible mechanism responsible
for this component: Violent SASI involves the development
of large-scale, large-amplitude density perturbations with
the same temporal dependence as the shock oscillations. 
Such
density perturbations $\delta \rho$ will directly contribute
to the signal through the term
$\rho x_i \partial_j \Phi$ in the integrand
of the quadrupole formula (\ref{eq:STFQ}) if the density
perturbations  $\delta \rho$ have an $l=2$ (quadrupole) component
$\delta \rho_2$.\footnote{Velocity
perturbations will, in principle, also contribute in Eq.~(\ref{eq:STFQ}).
Empirically, we find that their contribution to the GW amplitude
is minimal, however.}
Even $l=1$ sloshing and spiral motions will develop
a sizeable quadrupole component $\delta \rho_2$ in the
non-linear phase. The frequency of the emitted signal will trace the frequency of the underlying
SASI mode, but with frequency doubling for the $l=1$ mode
since the SASI-induced perturbation of the quadrupole moment will repeat itself after half a period as
the integrand in Eq.~(\ref{eq:STFQ}) is invariant
to a rotation by $\pi$ in any direction.
The contribution of the $l=1$ and $l=2$ modes
(and possibly their overtones) explains the double-peak
structure of the low-frequency signal in Figs.~\ref{fig:spectrograms},
\ref{fig:cuts} and \ref{fig:lowspec}. The amplitude
arising from the term $\rho x_i \partial_j \Phi$ will be of order
\begin{equation}
A 
\sim 
 \frac{G}{c^4} \int r\ \delta {\rho_2} \frac{\partial \Phi}{\partial r} \ud V
\sim
\frac{G}{c^4} \int \delta {\rho_2} \frac{G M}{r}  \ud V.
\end{equation}
The integral is essentially the potential energy stored in
$l=2$ density perturbations during SASI oscillations. Equating the
potential energy
with the kinetic energy in SASI motions and taking into account that
there is only a finite overlap with $l=2$, we find
\begin{equation}
A 
\lesssim 
\frac{G}{c^4} E_\mathrm{kin,SASI}.
\end{equation}
With $E_\mathrm{kin,SASI} \sim 10^{49} \, \mathrm{erg}$, we
obtain $A \lesssim 0.8 \, \mathrm{cm}$, which is roughly compatible
with the amplitudes (see Fig.~\ref{fig:lowpass}).

The anisotropic modulation of the accretion by the SASI is further
communicated to the PNS as material is advected downwards and settles
onto the PNS surface (something which may also be viewed as
non-resonant excitation of g-modes far below their
eigenfrequency). As matter seeps deeper into the outer layer of the PNS (layer~B) and
then even further down into the interior of the PNS (layer~A), it will still
emit GWs if the density and entropy perturbations are
not washed out completely by neutrino cooling. We have verified that
relatively large density fluctuations on the percent level are maintained
even in the cooling region. Since these density fluctuations still
retain a temporal modulation set by the SASI, they emit GWs in a similar, albeit somewhat broader frequency range. For the
same reasons as detailed above, the GWs amplitudes produced by
such a non-resonant excitation of g-modes will be related to
the kinetic energy stored in the mode and even a small kinetic
energy $\gtrsim 10^{48}  \, \mathrm{erg}$ in aspherical
mass motions below the gain region is sufficient to account
for the amplitudes.

The fact that the low-frequency signal from layer C
is \emph{weaker} than that from both layer A and layer B
is not in conflict with this explanation
because of cancellations in
the integral of $ \rho \left ( v_i v_j - x_i \partial_j \Phi \right) $
over the region outside the PNS, e.g.\ the overdensities in
the downflows can be compensated by the smaller shock radius
above them.\footnote{Immediately outside the minimum shock radius, the densities
of unshocked material above the downflows are \emph{lower} than in the shocked
material inside the high-entropy bubbles \emph{at a given radius}, i.e.\ overdensities
behind the shock correspond to underdensities at larger radii.}
Furthermore, we surmise that
density perturbations from the $l=1$ contribute more strongly to
the GW signal as they settle deeper into the PNS, because the pure
$l=1$ angular dependence of the perturbations in the post-shock
region develops a larger $l=2$ component during the process of settling.

The crucial role of the SASI in providing a slow, non-resonant forcing
of the outer regions of the PNS is also reflected in
the frequency structure of the signal. In Fig.~\ref{fig:sfreq} we
plot the Fourier amplitudes of the $l=1$ and $l=2$ components of the 
spherical harmonics decomposition of the shock position
for the period between $100 \, \mathrm{ms}$ and $350 \, \mathrm{ms}$ after bounce.
More precisely Fig.~\ref{fig:sfreq} shows
\begin{equation}
\sum_{m=-l,l}|\widetilde{a}_l^m(t)|^2 \ \ (l=1,2),
\end{equation} where $\widetilde{a}_l^m(t)$ is the Fourier transform of
\begin{equation} \label{eq:alsph}
a_l^m(t_n) = \frac{(-1)^{|m|}}{\sqrt{4\pi(2l+1)}} \int r_{\mathrm{sh}}(\theta,\phi,t) Y^m_l \ud \Omega.
\end{equation} Here, $r_{\mathrm{sh}}$ is the shock position (given by the Riemann-solver in our code) 
and $Y^m_l$ is the spherical harmonic of degree $l$ and order $m$.
Details about the shock can be found in \citet{hanke_13} for model s27, in \citet{hanke_phd} for models
s11.1 and s20  and in \citet{melson_15b} for model s20s.
The typical frequency for the $l=1$ mode ($50\ldots 100 \, \mathrm{Hz}$) and the
$l=2$ mode ($100 \ldots 160 \, \mathrm{Hz}$) of the shock are
compatible with the range of low-frequency emission seen in the GW
spectrograms, especially if we account for the fact that the GW signal
from forced $l=1$ motion will exhibit frequency doubling.

Since the
Fourier spectra of the $l=1$ and $l=2$ modes as well as the
GW spectrogram point towards a complicated frequency
structure with peak frequencies shifting in time (due to the variation
of the shock radius which sets the SASI frequencies) and contributions
from different phases interfering with each other in the
time-integrated spectrum, we refrain from a precise one-to-one
identification of the underlying modes.  

It is noteworthy that the effect of anisotropic accretion manifests
itself even down to the PNS convective layer. Apparently, the
eigenfunctions of the excited modes reach down quite deep through the
entire surface of the PNS (layer~B). However, the fact that even the deeper region of layer~A 
(below the overshooting region) contributes to
low-frequency GW emission suggests that $l=1$ and $l=2$ surface
motions can trigger convective motions (e.g.\ by providing density
perturbations that are then quickly amplified once they are advected
into the convectively unstable region).  Contrary to the mirror problem of
wave excitation at convective boundaries
\citep{goldreich_90,lecoanet_12}, such a coupling between the
accretion flow, the surface layer, and the PNS convection has as yet
been poorly explored. 

While the SASI is particularly effective at generating
a modulation of the accretion flow with a sizeable $l=2$ component,
large-scale convective motions in the hot-bubble region can also act as a substitute
for the SASI during periods of transient shock expansion (because
the typical scale of convective eddies is set
by the width of the unstable region, cp.\ \citealp{chandrasekhar_61,foglizzo_06}). The result
is a somewhat weaker and less sharply defined low-frequency
signal, which is what we observe during the
SASI-quiet periods in models s20s, s20, and s27 and also in model s11.2 (cf. Fig.~\ref{fig:spectrograms}).

With large-scale fluid motions in the gain region as the ultimate
agent responsible for low-frequency GW emission (through forced PNS
oscillations), the temporal structure of this signal component finds a
natural explanation. Generally, episodes of strong SASI activity
correlate with strong low-frequency GW activity. Large amplitudes of
the shock oscillations are not sufficient, however; the determining
factor is the kinetic energy contained in large-scale motions.  For
that reason, there is hardly any low-frequency emission
component during the second SASI episode in model s27.
During this phase, less mass is involved in SASI motions and
the SASI amplitude is significantly smaller. The lack of large-scale
motions with a significant $l=2$ component also explains the weak
low-frequency GW activity in model s11.2, where the
post-shock flow is dominated by smaller convective bubbles and the
kinetic energy in non-radial fluid motions is typically smaller than
for the more massive progenitors.

\subsection{Comparison of the exploding and non-exploding 20 solar mass models}
\begin{figure}
\includegraphics[width=0.49\textwidth]{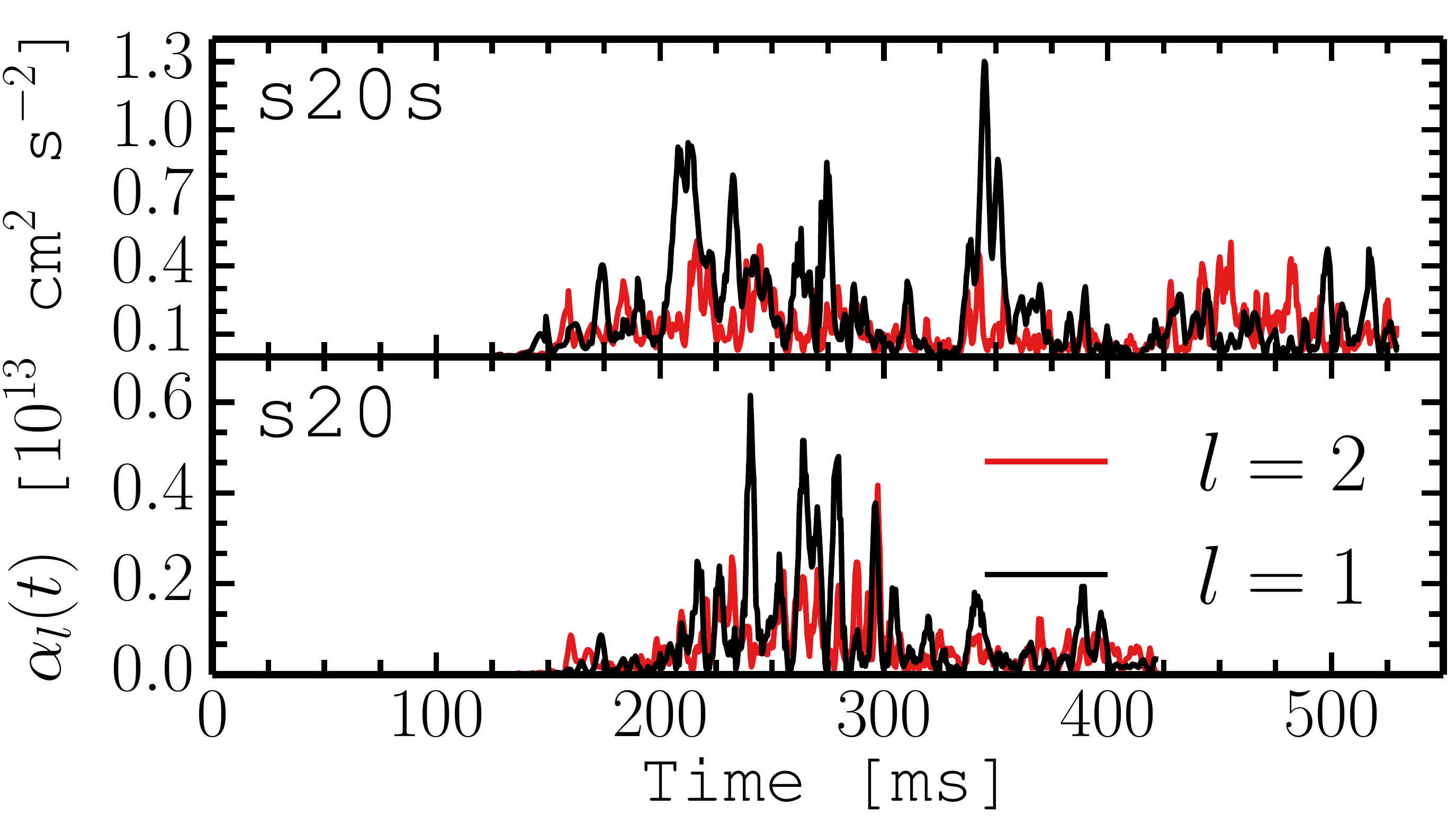}
\caption{The $l=2$ and $l=1$ components of the radial
velocity sampled in the unstable PNS layer for the $20 M_\odot$ models. The radial velocity has been sampled
at a radius $R$ given by $\rho(R) = 9.5 \times 10^{13}\, \mathrm{g} \, \mathrm{cm}^{-3}$.  The top panel
shows the exploding model s20s, and the bottom panel shows the non-exploding model s20, which was only calculated to $421 \, \mathrm{ms}$ after core bounce.
\label{fig:al2}
}
\end{figure}

The exploding model s20s differs only in details from its non-exploding
counterpart during the accretion phase. After the onset of shock expansion
strong low-frequency emission is sustained until the end of the simulation (see Fig.~\ref{fig:lowspec}).
This emission is connected to mass motions with a strong $l=2$ component in layer A.
In Fig.~\ref{fig:al2} we plot for models s20 and s20s,
\begin{equation}
\alpha_l = \sum_{m=-l}^{l}|\alpha_l^m(t)|^2 \ \ ,(l=1,2),
\end{equation} with
\begin{equation} \label{eq:alvr}
\alpha_l^m(t) = \frac{(-1)^{|m|}}{\sqrt{4\pi(2l+1)}} \int v_{r}(\theta,\phi,t) Y^m_l \ud \Omega,
\end{equation}
where $v_{r}$ is the radial velocity at a radius $R$ corresponding to an spherically averaged density of 
$\rho(R) = 9.5 \times 10^{13}\, \mathrm{g}\, \mathrm{cm}^{-3}$.

In the exploding model, the $l=2$ mode is generally stronger than in the non-exploding model and
it remains strong throughout the simulation in contrast to the non-exploding model,
where the $l=2$ mode decreases in strength after the SASI-dominated phase ends.
After a period of decreasing strength around 400 ms, the quadrupole mode in model s20s increases in 
strength and reaches amplitudes similar to those seen during the pre-explosion phase.
At the same time, there is a shift in the relative 
strength of the $l=1$ and $l=2$ mode after the onset of shock expansion,
While the quadrupole mode increases in strength, the dipole is relatively weak at late times.
This transition into a flow pattern that is dominated by an $l=2$ mode resonates better with the
quadrupole nature of GW emission. We therefore see an increase in low-frequency emission
from the unstable layer within the PNS.
Such a change in the spectrum of eddy scales after shock revival
could result from changes in the asymmetric accretion flow onto
the PNS, or from changes in the stratification of entropy and electron
fraction, but for the purpose of interpreting the GW
emission, the ultimate reason is immaterial and left to more detailed future studies of the
hydrodynamics of PNS convection.
Aliasing of high-frequency emission may also be partly responsible for the enhanced low-frequency emission after shock revival. However,
since the spectrograms in Fig.~\ref{fig:spectrograms} show broadband emission, it is unlikely that the low-frequency emission we see 
after shock revival is caused by aliasing effects alone. 

\section{Detection Prospects} \label{sec:obs}
\begin{table*}
\centering
\caption{Signal-to-noise ratio (SNR) for all four models. Values
are given for two different frequency domains, $20\ldots 250 \, \mathrm{Hz}$
(low-frequency) and $250\ldots 1200 \, \mathrm{Hz}$
(high-frequency). The table shows values for two different
detectors, AdvLIGO and the Einstein Telescope. For the latter
we calculate the SNR for two
different modes of operation (ET-B and ET-C).  SNRs have been
computed for a source at a distance of $10 \, \mathrm{kpc}$. 
For model s20s we only show the
SNR for the low-frequency band since the high-frequency band 
is somewhat contaminated
by aliasing effects. For s27, s20, and s11.2 we also give the ratio of the band-limited SNRs in the low- and high-frequency bands to quantify
the ``colour'' of the GW signal.
\label{table:SNR}
}
\begin{tabular}{l|l|l|l|l|l|l|l|l|l|l|l|l|l|l|l|l|}
\cline{2-16}
\multirow{2}{*}{}             & \multicolumn{4}{c|}{s27}               & \multicolumn{4}{c|}{s20}                                  & \multicolumn{3}{c|}{s20s}  & \multicolumn{4}{c|}{s11.2}  \\ \cline{2-16} 
                              & Low      & High    & Total  & Low/High &      Low      & High     & Total  & Low/High & Low      & High    & Total & Low      & High    & Total   & Low/High       \\ \hline
\multicolumn{1}{|l|}{AdvLIGO} & 3.7      & 4.5     & 8.8    &  0.82    & 5.3           & 7.7      & 9.4   &   0.82    & 10.2     & --      & --    & 1.3      & 4.1     & 4.3    &    0.32      \\ \hline
\multicolumn{1}{|l|}{ET-C}    & 50.0     & 64.0    & 81.3   &  0.78    & 73.9          & 109.3    & 131.9 &   0.83    & 139.7    & --      & --    & 18.1     & 50.9    & 53.9   &    0.36      \\ \hline
\multicolumn{1}{|l|}{ET-B}    & 78.5     & 73.7    & 107.7  &  1.07    & 113.9          & 127.0    & 170.6  &   0.74  & 217.3    & --      & --    & 28.0     & 67.3    & 72.8   &    0.42       \\ \hline
\end{tabular}
\end{table*}

With the prominent high-frequency component of the signal in 2D
largely muted in 3D, it is evidently necessary to reconsider the
prospects of detecting a Galactic supernova. Detailed detectability
studies based on 2D waveforms \citep{logue_12} may now well be too
optimistic after the update of the waveform predictions.  While an
elaborate statistical machinery is required to reliably determine
signal detectability and possible inferences about core-collapse
supernova physics \citep{logue_12,hayama_15}, we can
already draw some conclusions for the waveforms presented
in this paper.

\subsection{General Considerations}
The detectability of the GW signal from a core-collapse supernova has
often been assessed using the signal-to-noise ratio (SNR) for matched
filtering. Assuming an optimally-orientated detector and a roughly
isotropic frequency spectrum for different observer directions, the
SNR for matched filtering is formally given by \citep[][cp.\ their Eq.~(5.2) for the second form]{flanagan_98a},
\begin{equation}
\label{eq:snr}
(\mathrm{SNR})^2
=
4 \int_0^\infty df \frac{|\tilde{h}(f)|^2}{S(f)}
=
\int_0^{\infty} df \frac{h_c ^2}{f^2 S(f)},
\end{equation}
where
\begin{equation}
\label{eq:h_c}
h_c =\sqrt{ \frac{2 G}{\pi^2 c^3 D^2} \frac{d E_\mathrm{GW}}{df}}
\end{equation}
is the characteristic strain, $S(f)$ is the power-spectral density of the detector noise as a
function of frequency $f$, and $d E_{GW}/df$ is the spectral energy density of the GWs.
Note that the second expression for the SNR in Eq.~(\ref{eq:snr})
has been obtained under the assumption of isotropic GW emission so that
one can express the (formally) direction-dependent squared
amplitudes in terms of the GW energy spectrum $d E_\mathrm{GW}/df$.

Since the GW signal of a core-collapse supernova is,
however, \emph{not amenable to matched filtering} because of its
stochastic character, the SNR formally defined by
Eq.~(\ref{eq:snr}) must be interpreted with care. 

The SNR still remains a
useful quantity as it measures the excess power during the time of
integration, as can be seen by re-expressing Eq.~(\ref{eq:snr}) in
terms of the expectation value for the Fourier coefficients
$\tilde{n}(f)$ of the noise over a finite time-interval $\Delta t$
(the integration time for the signal),
which obey (cp.\ \citealp{logue_12})
\begin{equation}
\langle\tilde{n}(f)\tilde{n}^*(f) \Delta f\rangle=S(f)/2,
\end{equation}
where the factor $1/2$ appears because $S(f)$ is defined as the one-sided power
spectral density of the time-dependent strain noise $n(t)$.
Note that the frequency spacing $\Delta f$ is given by $\Delta f=
1/\Delta t$ and that $\Delta t$ can be set to the length of the
signal in consideration in our case.
For a finite time series, where the integral in 
Eq.~(\ref{eq:snr}) can be replaced with a sum over the Fourier modes
at discrete frequencies $f_k=k/\Delta t$ (with integer k), we then obtain,
\begin{equation}
(\mathrm{SNR})^2=
8\sum_k \frac{|\tilde{h}(f_k)|^2 \Delta f}{\langle |\tilde{n}(f_k)|^2\rangle \Delta f}
=8\sum_k \frac{|\tilde{h}(f_k)|^2 }{\langle |\tilde{n}(f_k)|^2\rangle}.
\end{equation}
For uncorrelated Gaussian noise in each frequency bin, the SNR of a
prospective signal obtained from the summation over $N_\mathrm{bins}$
frequency bins is thus related to the $\chi^2$-value for this signal
as
\begin{equation}
\chi^2\sim N_\mathrm{bins}+\mathrm{SNR}^2/2,
\end{equation}
where the additional term $N_\mathrm{bins}$ comes from the
contribution of the noise in each bin. Sufficiently high values of
$\chi$ during a prospective supernova event (with an integration
interval  $\Delta t$ defined by a coincident neutrino signal)\footnote{This is
  crucial because it is always possible to find short intervals with
  power excess comparable to a physical signal if the integration time
  is sufficiently long.} can be attributed to a physical signal;
e.g.\ to exclude stochastic fluctuations as a source of the excess
power at a confidence level of $95\%$, one needs
\begin{equation}
\label{eq:snr_band1} 
\mathrm{SNR}^2/8=\chi^2-N_\mathrm{bins} \gtrsim 2.3 \sqrt{N_\mathrm{bins}}
\end{equation}
for large $N_\mathrm{bins}$. For a signal with power excess
in a frequency band with bandwidth $\delta f$ and $N_\mathrm{bins}=\delta f/\Delta f= \delta f\, \Delta t$,
this implies the requirement
\begin{equation}
\label{eq:snr_band2}
\mathrm{SNR} \gtrsim 4.3 \, (\delta f\, \Delta t)^{1/4},
\end{equation}
for a detection of a signal in this band. This roughly corresponds to
the results obtained by \citet{flanagan_98a} for noise monitoring in
Section~IIB of their paper.

Prior knowledge of the signal structure
can help to identify signals  with even lower SNR; \citet{logue_12},
for example, showed that a detection and identification can be
possible already for $\mathrm{SNR} \sim 10$ with the help of a
principal component analysis of template waveforms provided that the
signal structure is not too dissimilar from the template. This is
in line with the weak dependence of Eq.~(\ref{eq:snr_band2})
on the bandwidth $\delta f$.

If properly interpreted, the SNR thus remains a useful measure for the
detectability of our predicted signals within the scope of this paper.
Its inherent limitations provide justification for neglecting the
effect of the detector orientation and the
precise directional dependence of the signal by computing the SNR from
the energy spectrum $d E/df$ instead of a direction-dependent Fourier
spectra of the strain. We have verified that the SNR for
the low-frequency band does not vary by more than $\mathord{\sim} 20\%$,
and the SNR for the high-frequency signal depends even less on the
observer direction.

\subsection{Detection Prospects for Simulated Models}
We calculate the SNR from Eq.~(\ref{eq:snr}) for the zero-detuning-high power configuration
of Advanced LIGO \citep{adv_sens} and the B \citep{et_b} and C \citep{et_c} configurations for the Einstein
telescope. We refer to these configurations as AdvLIGO, ET-B
and ET-C. In order to better assess the detectability and possible
inferences from the signal structure, we compute SNRs quantifying the
excess power in a low-frequency band ($\mathrm{SNR}_\mathrm{low}$ for
$20 \, \mathrm{Hz} \le f < 250 \, \mathrm{Hz}$,
i.e.\ $\delta f= 230 \, \mathrm{Hz}$) and a high-frequency
band ($\mathrm{SNR}_\mathrm{high}$ for $250 \, \mathrm{Hz} \le f <
1200 \, \mathrm{Hz}$, i.e.\ $\delta f= 950 \, \mathrm{Hz}$). SNRs for all models in
those two bands for events at a distance of $10 \, \mathrm{kpc}$ are
presented in Table~\ref{table:SNR}. Using
Eq.~(\ref{eq:snr_band2}), we obtain a detection threshold of
$\mathrm{SNR}_{\mathrm{low}} \gtrsim 11$ for the low-frequency band and
$\mathrm{SNR}_{\mathrm{high}} \gtrsim 15$ for the high-frequency band assuming
$\Delta t=0.5 \, \mathrm{s}$. Since the critical SNR depends
weakly on $\Delta t$, these fiducial values can be used for all
models. SNRs for arbitrary distances can easily be obtained since the
SNR is inversely proportional to the distance.

Regardless of the precise detector configuration, the SASI-dominated
models s20 and s27 are clearly distinguished from the
convective model s11.2 through a higher ratio
$\mathrm{SNR}_\mathrm{low}/\mathrm{SNR}_\mathrm{high} >0.65$ compared
to $\mathrm{SNR}_\mathrm{low}/\mathrm{SNR}_\mathrm{high} <0.42$.
SASI-dominated models thus appear ``redder'' in GWs
before the onset of the explosion.
Based on our small sample, they also appear to be characterised by a higher SNR,
but this might be incidental. More massive progenitors with
stronger neutrino heating in the gain region, stronger cooling above
the PNS convection, and a larger mass in the gain region could 
produce a stronger GW signal, even in the absence of strong SASI activity. The ratio
$\mathrm{SNR}_\mathrm{low}/\mathrm{SNR}_\mathrm{high}$, on the other hand, should be a robust indicator for the presence or absence of
large-scale SASI motions.

Note that since model s20s suffers most severely from aliasing effects, the SNR
in the high-frequency domain might be inaccurate. We therefore refrain from giving values for $\mathrm{SNR}_{\mathrm{high}}$ and the total SNR.  The low-frequency band, on the other hand, should be unaffected by aliasing artefacts and $\mathrm{SNR}_\mathrm{low}$ is significantly
higher than in the non-exploding models. It is possible that the
enhanced low-frequency emission from the convectively unstable region
of the PNS is a general feature in exploding models and we
hypothesise that shock revival will be followed by GW emission with \emph{excess power in
  the low-frequency band}. This is in contrast to previous studies in 2D
\citep{murphy_09,mueller_13} where shock expansion is typically
followed by an \emph{increase in the high-frequency emission band}.
If shock revival generally leads to enhanced low-frequency emission,
this would obviously complicate the interpretation of
a high value of $\mathrm{SNR}_\mathrm{high}/\mathrm{SNR}_\mathrm{low}$,
which could \emph{either} indicate SASI activity or the transition
to an explosion.

\subsection{Detection Prospects with AdvLIGO}
For a supernova at a distance of $10 \, \mathrm{kpc}$, it is evident
that \emph{none} of the four models could be detected by AdvLIGO
based on excess signal power. Given the reduction of the typical
amplitudes by a factor of $\mathord{\sim} 10$ in 3D compared to 2D,
this is not surprising. Using an approach based on simulated noise and
a principal component analysis of the signal, \citet{logue_12}
and \citet{gossan_15} already
found that AdvLIGO is only marginally able to identify waveforms
from 2D supernova simulations for events at distances of a few
$\mathrm{kpc}$. 

For the SASI-dominated models (s20, s27, s20s), the excess
power in the low-frequency band would become detectable
at $95 \%$ confidence level at the distance of the Crab supernova
($\mathord{\sim} 2 \, \mathrm{kpc}$), as would the high-frequency
component of model s20. Model s11.2, on the other hand, would not
show a statistically significant
power excess. 

\subsection{Detection Prospects with the Einstein Telescope}
The situation will change drastically with the Einstein Telescope.  For
either configuration considered here, the excess power in both bands
ought to be detectable for an event at a distance of $10
\, \mathrm{kpc}$, although the low-frequency component of model s11.2
would barely make it above the detection threshold
for ET-C. The high SNR in both bands would permit
a measurement of $\mathrm{SNR}_\mathrm{low}/\mathrm{SNR}_\mathrm{high}$
as an indicator for the GW ``colour'' with some confidence.
Even at a distance of $20 \, \mathrm{kpc}$, the excess power in
both the high- and low-frequency bands would still remain detectable
and quantifiable in the SASI-dominated models.
For the more modest goal of a mere detection, the SNR for
model s20s would be high enough to observe events throughout
the entire Milky Way and even out to the Large Magellanic
Cloud ($\mathord{\sim} 50 \, \mathrm{kpc}$).

\subsection{Interpretation of a Prospective Detection}
Without a more sophisticated analysis of the time-frequency
structure of a prospective detection event, only limited conclusions
about the supernova core could be drawn from excess power
measured by GW detectors during specific time windows. Nonetheless, a GW
detection with the Einstein Telescope would be valuable for corroborating
our understanding of hydrodynamic instabilities in the core
in conjunction with the observed neutrino signal.

A high value of
$\mathrm{SNR}_\mathrm{low}/\mathrm{SNR}_\mathrm{high}$ concurrent
with a periodic modulation of the neutrino signal
\citep{marek_08,lund_10,brandt_11,tamborra_13, mueller_14,tamborra_14b} would
furnish solid evidence for SASI activity, and strong
low-frequency emission concurrent with modulations of the
neutrino signal below $\mathord{\sim} 50 \, \mathrm{Hz}$ would strongly
indicate that shock revival is already underway during the
time window in question. While these conclusions could likely
be drawn on the basis of the neutrino signal alone for nearby
supernovae with a suitable orientation of the SASI spiral
plane or sloshing mode, the detection of modulations
in the neutrino signal for non-optimal orientations becomes
difficult at
distances $\gtrsim 10 \, \mathrm{kpc}$
\citep{mueller_14}. In such cases, combining the GW and neutrino
signal would likely allow stronger conclusions.

When SASI-induced modulations of the neutrino signal are not detectable due to
distance, orientation, or unfavourable neutrino flavor oscillations, a detection of strong GW power in the
low-frequency band would still provide evidence for \emph{either}
SASI activity (since this signal component is more robust against
orientation effects than modulations of the neutrino signal)
\emph{or} the onset of strongly asymmetric accretion after shock
revival. If the SNR is sufficiently high to localise the GW power excess in time relative to the onset of the neutrino signal (which roughly marks
the time of bounce), it may be possible to decide between those two alternatives.

Late GW power excess after $\mathord{\gtrsim} 0.5 \, \mathrm{s}$ will
likely indicate the onset of the explosion without prior SASI activity,
since the SASI typically reaches non-linear saturation well before this point,
and since the decreasing mass in the gain region does not allow for strong
late-time GW emission due to the SASI (as shown by models s20 and s27).

\section{Conclusions} \label{sec:con}
We have studied the GW signal from the accretion phase and the early
explosion phase of core-collapse supernovae based on four recent 3D
multi-group neutrino hydrodynamics simulations. We considered four models based on three
progenitors with ZAMS masses of $11.2 M_\odot$, $20 M_\odot$, and $27 M_\odot$.
The three non-exploding models enabled us to study the phase between
bounce and shock revival. We covered both the SASI-dominated regime
(model s20, \citealp{tamborra_14b}; model s27, \citealp{hanke_13}), 
as well as the   
convection-dominated regime (model s11.2, \citealp{tamborra_14a}).
Additionally, the exploding $20 M_\odot$ model s20s
\citep[][with a modified axial-vector coupling constant for neutral
  current scattering]{melson_15b} illustrates changes in the GW signal
in exploding models.  Since our treatment of the microphysics and the
neutrino transport is on par with previous works on the GW signal from
2D simulations \citep{marek_08,yakunin_10,mueller_13,yakunin_15}, we
were in the position to conduct a meaningful comparison of GW emission
in 2D and 3D during the accretion and explosion phase for the first
time. To this end, we included the $27 M_\odot$ 2D models of
\citet{mueller_12b} and  \citet{hanke_13} in our study.

Our analysis showed differences between the GW emission in 2D and
3D. The prominent, relatively narrow-banded emission at
high-frequencies that is characteristic of 2D models is significantly
reduced. With the reduction of the high-frequency emission,
distinctive broadband \emph{low-frequency} emission in the range between $100 \, \mathrm{Hz}$
and $200 \, \mathrm{Hz}$ emerges as a characteristic feature during
episodes of SASI activity and during the explosion phase of model s20s.
The low-frequency emission does also exist in the 2D models, but it is completely
overwhelmed by the high-frequency emission. 
This conclusion is somewhat model dependent, because in 
one of our 2D models, s27-2D, high-frequency GW emission
is low and the low-frequency component becomes very prominent.

We discussed these differences extensively from two vantage points:
On the one hand, we investigated the underlying hydrodynamic processes
responsible for GW emission and showed how the changes in the GW signal in 3D
are related to critical differences in flow dynamics in 3D compared to 2D.
On the other hand, we outlined the repercussions of these changes
for future GW observations and sketched possible inferences that
could be drawn from the detection of a Galactic event by third-generation
instruments.

With regard to the hydrodynamic processes responsible for GW
emission, our findings can be summarised as follows:
\begin{enumerate}
\item There is a high-frequency signal component that closely traces
  the buoyancy frequency in the PNS surface region in 2D and 3D,
  i.e.\ the roughly isothermal atmosphere layer between the PNS
  convection zone and the gain region acts as frequency stabiliser for
  forced oscillatory motions in both cases.  However, the
  high-frequency component mostly stems from aspherical mass motions
  in and close to the overshooting region of PNS convection in 3D,
  whereas it stems from mass motion close to the gain radius in
  2D. This indicates that quasi-oscillatory mass motions at
  high frequencies are instigated \emph{only by PNS
    convection in 3D} even during the pre-explosion phase, whereas
  forcing by the SASI and convection in the gain region is dominant in
  2D. The resulting \emph{amplitudes of the high-frequency component
  are considerably lower in 3D than in 2D}.
\item We ascribe the strong excitation of high-frequency surface
  g-mode oscillations in 2D to several causes: The inverse turbulent
  cascade in 2D leads to larger impact velocities of the downflows and
  creates large flow structures that can effectively excite $l=2$
  oscillations that give rise to GW emission. Braking of downflows by
  the forward turbulent cascade and fragmentation into smaller eddies strongly
  suppress surface g-mode excitation in 3D. Moreover, the spectrum of
  turbulent motions does not extend to high frequencies in 3D both in
  SASI-dominated and convection-dominated models so that the resonant
  excitation of the $l=2$ surface g-mode at its eigenfrequency becomes
  ineffective.
\item In 3D, low-frequency GW emission in the pre-explosion phase
  ultimately stems from the global modulation of the accretion flow by
  the SASI. Because of frequency doubling and/or the contribution
  from the $l=2$ mode, the typical frequencies of this component are
  of the order of $100 \ldots 200 \, \mathrm{Hz}$, i.e.\ somewhat
  higher than the typical frequency of the $l=1$ modes of the SASI.
  Mass motions in the post-shock region, the PNS surface
  region and the PNS convection zone all contribute to this
  low-frequency component, which indicates that the modulation of the
  accretion flow is still felt deep below the gain radius as the accreted
  matter settles down onto the PNS.
  Moreover, our analysis of the detection prospects shows that 
\emph{the low-frequency component of the signal at $\mathord{\gtrsim} 100 \, \mathrm{Hz}$
 becomes a primary target in terms of detectability} in contrast to previous 2D results.     
\item By contrast, convective models characterised by mass motions of
  intermediate- and small-scale like s11.2 show very little GW emission at low frequencies. 
  The high-frequency emission, on the other hand, is excited primarily by PNS convection and is therefore less sensitive to the dominant 
  instability (convection or SASI) in the post-shock region.
  \emph{Thus, the ratio of high-frequency to low-frequency GW power
  can potentially be used to distinguish SASI- and convection-dominated
  models in the pre-explosion phase.}
\item However, strongly enhanced low-frequency emission can also occur
  due to a change of the preferred scale of the convective eddies in
  the PNS convection zone as exemplified by model s20s, where the
  dominant mode shifts from $l=1$ to $l=2$ late in the
  simulation. Since this does not occur in the corresponding
  non-exploding model s20, one can speculate that this behaviour is due
  to changes in the accretion flow and neutron star cooling associated
  with shock revival. If this behaviour is generic for exploding
  models enhanced GW emission may still remain a fingerprint
  of shock revival as it is in 2D \citep{murphy_09,mueller_13}. With
  only one explosion model available to us, this conclusion does not
  rest on safe ground; more 3D explosion models are needed to check
  whether enhanced low-frequency GW emission after shock revival is
  indeed a generic phenomenon.
\end{enumerate}

It is obviously of interest whether future GW observations will be
able to discriminate between models with such distinctively different
behaviour as the ones presented here. Without an elaborate statistical
analysis, only limited conclusions can be drawn concerning this point. In this
paper, we confined ourselves to rough estimates based on the expected
excess power in second- and third-generation GW detectors in two bands
at low ($20 \ldots 250 \, \mathrm{Hz}$) and high ($250 \ldots 1200
\, \mathrm{Hz}$) frequency. Due to the reduction of the signal
amplitudes compared to 3D, the prospects for second-generation
detectors appear rather bleak; even the SASI-dominated models s20,
s20s, and s27 could not be detected out further than $\mathord{\sim} 2
\, \mathrm{kpc}$ with AdvLIGO at a confidence level of
$95\%$. Third-generation instruments like the Einstein Telescope,
however, could not only detect all of our models at the typical
distance of a Galactic supernova ($\mathord{\sim} 10 \, \mathrm{kpc}$) and
strong GW emitters like s20s out to $50 \, \mathrm{kpc}$; the expected
signal-to-noise ratios could even be high enough to distinguish models
with enhanced low-frequency emission due to SASI from convective
models based on the ``colour'' of the GW spectrum. In conjunction with
timing information and the neutrino signal, it may also be possible
to distinguish enhanced low-frequency emission from the SASI
from enhanced GW emission after shock revival as in model s20s.

However, more work is obviously needed to fully exploit the potential
of GWs as a probe of the supernova engine in the case of ``ordinary'',
slowly rotating supernovae for which PNS convection and the SASI are
the dominant sources of GW emission. Desiderata for
the future include a much broader range of 3D explosion
models to determine to what extent the aforementioned features in the
GW signal are generic. With waveforms from longer explosion simulations,
the prospects for detecting a Galactic supernova in GWs with second
generation instruments may also
appear less bleak than they do now based on our biased selection that
includes only one explosion model evolved to $200 \, \mathrm{ms}$
after shock revival.

Furthermore, it is conceivable that much more information can be
harvested from the GW signals than our simple analysis
suggests. Several authors \citep{logue_12,hayama_15,gossan_15} have
already demonstrated the usefulness of a powerful statistical
machinery in assessing the detectability of supernovae in GWs and
distinguishing different waveforms (e.g.\ from rotational collapse and
hot-bubble convection, \citealp{logue_12}). Peeling out the more
subtle differences between SASI- and convection dominated models from
GW signals in the face of greatly reduced signal amplitudes certainly
presents a greater challenge, but third-generation instruments will 
nonetheless make it an effort worth undertaking.

The GW analysis presented in this work is based on three non-rotating
progenitors, {\comment and it remains to be seen whether the findings
  from these simulations are generic. For GW detection, it is
  particularly important to ascertain whether the overall reduction of
  the signal from SASI and convection in 3D compared to 2D is always
  as strong as in our models, where the difference is a factor of
  $\mathord{\sim} 10$. This has recently been questioned by
  \citet{yakunin_17}, who reported considerably higher
  amplitudes for a $15 M_\odot$ progenitor than in our models and
  found the energy emitted in GW to be similar in their 2D and 3D
  simulations. Considering that we obtain weaker GW signals in 3D in
  models that probe a variety of different regimes, and that other 3D
  studies \citep{mueller_e_12,kuroda_16} predict amplitudes in line
  with our findings (albeit with less rigorous neutrino transport and
  without a 2D/3D comparison) suggests that small amplitudes
  $|A|\lesssim 5 \, \mathrm{cm}$ are generic in 3D and that the strong
  amplitudes in \citet{yakunin_17} are the exception rather than the
  norm and need further explanation. Nonetheless, the range of variation in GW amplitude in 3D
deserves to be explored further in the future.

There are various properties of the pre-collapse cores that will
(or at least could) impact the GW signal. The influence of rotation is well known:}
 In rapidly rotating models there is a strong GW burst
associated with the rebound of the core \citep{mueller_82}.
During the post-bounce phase rotation can lead to a
bar-like deformation of the core \citep{rampp_98,shibata_05} or the development
of low-mode spiral instabilities \citep{ott_05,kuroda_14,takiwaki_16}.
These flow patterns in turn lead to strong GW emission at frequencies 
determined by the rotational frequency.
In addition, rotation can modulate processes
already present in nonrotating models, for example prompt convection or the SASI. 
In the models presented by \citet{dimmelmeier_08} and \citet{ott_12}
only models with moderate rotation rates (and nonrotating models) exhibit prompt convection.
The coupling between rotation and SASI activity can lead to an enhanced growth 
rate of the spiral SASI mode \citep{blondin_07a,yamasaki_08,iwakami_09,kazeroni_16,janka_16}. 
Whether a significant proportion of supernova progenitors have moderately rotating (let alone rapidly rotating) cores is unclear. 
Stellar evolution models that include the effects of magnetic fields predict rather slowly rotating
pre-collapse cores \citep{heger_05}. Furthermore, the angular momentum loss due to stellar winds 
seems to be underestimated by stellar evolution models, compared to results from asteroseismology \citep{cantiello_14}.
Predictions of the initial rotation rate of pulsars, based on their current
spin-down rate and age, suggest that a large fraction of the pulsar population is born 
with rotation periods of the order of tens to hundreds of milliseconds \citep{popov_12,noutsos_13}. 

There is also the issue of starting the simulations from spherically symmetric progenitor models.
{\comment Current GW predictions like ours rely on explicitly imposed (this study) or numerical
seed perturbations to trigger the development of non-radial instabilities, and it needs
to be explored further whether the level of seed perturbations is partly responsible for
differences in the GW amplitudes calculated by different groups (e.g.\ this study and \citealt{yakunin_17}).
Moreover, it} has been found that {\comment physical seed} asymmetries in the burning shells of the progenitor can 
influence the shock dynamics and even help to ensure a
successful explosion \citep{burrows_96,fryer_04,arnett_11,couch_13,mueller_15a}. Any change
in the initial conditions that leads to a significant change in the dynamics of the
supernova core should be expected to impact the GW signal. Therefore, it will 
be important to keep improving the predicted GW signals, in hand with the improvement of
core collapse models.

\section{Acknowledgements}
We thank Maxime Viallet, Jerome Guilet, Gerhard Sch\"afer, Paul Lasky, and Eric Thrane for useful discussions. 
At Garching, this work has been supported by the Deutsche Forschungsgemeinschaft through the Excellence Cluster Universe EXC
153 and by the European Research Council through
ERC-AdG No. 341157-COCO2CASA. 
Bernhard M\"uller acknowledges support by the Australian Research
Council through a Discovery Early Career Researcher Award (grant
DE150101145).  The simulations
were performed using
high-performance computing resources (Tier-0) provided
by PRACE on CURIE TN (GENCI@CEA, France) and
SuperMUC (GCS@LRZ, Germany).

\bibliography{paper}

\end{document}